\RequirePackage{fix-cm}
\documentclass[smallextended]{svjour3}       % onecolumn (second format)
\smartqed  % flush right qed marks, e.g. at end of proof
\usepackage{graphicx}
\usepackage{booktabs}
\usepackage{cite}
\usepackage{paralist}
\usepackage{longtable}
\usepackage{lscape}
\usepackage[titletoc,toc,title]{appendix}

\usepackage{verbatim} %for multiline comments
\usepackage{textcomp}
\usepackage{longtable} % multi-page tables

\begin{document}

% \title{What to Share, When and Where?}
% \subtitle{A multiple-case study on the rationale behind open source}
% \title{What to Share as Open Source Software: Balancing Objectives and Complexities}
\title{What to Share, When, and Where: Balancing the Objectives and Complexities of Open Source Software Contributions}

\titlerunning{What to Share, When, and Where}        % if too long for running head

\author{Johan~Lin{\aa}ker \and
        Bj\"{o}rn~Regnell
}


\institute{J. Lin{\aa}ker \at
              Box 118, SE-221 00 Lund \\
              Tel.: +46 46 222 49 27\\
              Fax: +46 46 13 10 21 \\
              \email{johan.linaker@cs.lth.se}           %  \\
           \and
           B. Regnell \at
              \email{bjorn.regnell@cs.lth.se}
}

\date{Received: date / Accepted: date}

% This study aims to empirically investigate objectives and complexities organizations need to consider and balance when deciding on what software to share as OSS, when to share it, and whether to create new or use existing community

\maketitle

\begin{abstract}
\textbf{Context:} Software-intensive organizations' rationale for sharing Open Source Software (OSS) may be driven by both idealistic, strategic and commercial objectives, and include both monetary as well as non-monetary benefits. To gain the potential benefits, an organization may need to consider what they share and how, while taking into account risks, costs and other complexities.
\textbf{Objective:} This study aims to empirically investigate objectives and complexities organizations need to consider and balance between when deciding on what software to share as OSS, when to share it, and whether to create a new or contribute to an existing community.
\textbf{Method:} A multiple-case study of three case organizations was conducted in two research cycles, with data gathered from interviews with 20 practitioners from these organizations. The data was analyzed qualitatively in an inductive and iterative coding process. 
\textbf{Results:} 12 contribution objectives and 15 contribution complexities were found. Objectives include opportunities for improving reputation, managing suppliers, managing partners and competitors, and exploiting externally available knowledge and resources. Complexities include risk of loosing control, risk of giving away competitive advantage, risk of creating negative exposure, costs of contributing, and the possibility and need to contribute to an existing or new community.
\textbf{Conclusions:} Cross-case analysis and interview validation show that the identified objectives and complexities offer organizations a possibility to reflect on and adapt their contribution strategies based on their specific contexts and business goals.

\end{abstract}

\section{Introduction}\label{sec:introduction}
\label{intro}

Sharing, or contributing, software artifacts (e.g., features, projects, and frameworks) as Open Source Software (OSS) is a common practice among software-intensive organizations today~\cite{munir2015open}. By ''opening up''~\cite{chesbrough2014new}, an organization can exploit the external workforce residing in the OSS communities that develops and maintains the OSS. Improved product innovation, accelerated development, lower maintenance cost, as well as improved branding and reputation, are some of the potential benefits that may motivate~\cite{munir2015open, munir2017open, stuermer2009extending, henkel2006selective, lindman2009beyond}. The motive may also be driven by pure idealism and being a good OSS citizen~\cite{jansen2012shades}. 
For some organizations, OSS may have a more direct connection to the business model or strategy, e.g., as a basis for complementary products and services~\cite{sharivar2018business, andersen2012commercial, spijkerman2018open}, or as a means to create a new standard or compete with existing ones~\cite{lindman2009beyond, jansen2012shades, henkel2006selective, west2003open}. Objectives for why an organization would choose to contribute software artifacts as OSS does, however, not have to be limited to one or the other~\cite{andersen2012commercial}. 

In this paper, we present the results from an empirical study on organizations' rationale for sharing software artifacts as open source. In the context of this study, we introduce the term \textbf{Contribution Objective} (CO), which we define as \textit{a purpose for contributing a software artifact, motivated by a monetary or non-monetary benefit that is enabled or resulted directly or indirectly as a consequence of the contribution}.

To gain potential benefits by acting in line with contribution objectives, an organization needs to access the external workforce, either by contributing its software artifacts to an existing OSS community or by creating a new community, each with its respective costs and risks~\cite{dahlander2008firms}. In either case, the organization then needs to work actively to align its internal strategy with the community where they are a stakeholder among many, potentially including competitors with conflicting agendas~\cite{dahlander2005relationships, dahlander2006man}. An organization, therefore, may have to consider not just \emph{where}, but also \emph{what} it contributes, and \emph{when}. Risks include giving away differentiating functionality~\cite{wnuk2012can,van2009commodification, henkel2006selective,henkel2008champions,west2006challenges,iivari2008usability}, or contributing too late and having to choose between adopting the alternative solution or maintaining an internal solution alone~\cite{linaaker2018motivating, wnuk2012can}. Hence, there are several potential costs and risks tied to a contribution. 

In the context of this study, we refer to these potential costs and risks as \textbf{Contribution Complexities} (CCs) and define them as \textit{aspects related to a software artifact that may complicate the contribution of the artifact, or imply a cost or risk as a result thereof, either directly or indirectly}. As with contribution objectives, not all complexities may be relevant for all organizations~\cite{linaaker2018motivating}.

By not considering relevant contribution objectives or complexities, an organization may risk making a contribution that could be damaging, or inadvertently block a contribution that could have been beneficial for the organization and its business goals~\cite{wnuk2012can}. To gain the expected benefits, organizations therefore need to link their business goals with their decisions on what they contribute as OSS. Despite the problematic context, research on how organizations can develop and use such strategies is limited~\cite{munir2015open}, with some exceptions~\cite{wnuk2012can, weikert2013model, linaaker2018motivating}. This leads us to define the following research questions:

\begin{itemize}
    \item[\textbf{RQ1}] What contribution \textit{objectives} should a software-intensive organization consider when assessing if, where and when a software artifact should be shared as OSS?
\end{itemize}

\begin{itemize}
    \item[\textbf{RQ2}] What contribution \textit{complexities} should a software-intensive organization consider when assessing if, where and when a software artifact should be shared as OSS?
\end{itemize}

We addressed these research questions by conducting a multiple-case study at three software-intensive organizations with an iterative approach spanning over two research cycles. Based on an inductive coding of twenty semi-structured interviews divided among the three organizations, 12 contribution objectives and 15 contribution complexities were identified.

An organization may, based on the contribution objectives and complexities that they find relevant for their context and business goals, make informed decisions on what software artifacts that should be released as OSS, when and where. For a specific artifact, the question ``\textit{what?}'' regards if the artifact should be contributed in full or kept closed, or if certain parts can be contributed under certain conditions. The question ``\textit{when?}'' refers to when in time an artifact should be contributed. Finally, the question ``\textit{where?}'' asks whether the artifact should be contributed to one of many existing OSS communities or if a new community should be established. Answers to these questions are input to a \textit{contribution strategy} for the software artifact under consideration. The overall objective of the work presented in this paper is to elicit such answers from real-world example cases, and start building a relevant list of considerations that can help when developing a contribution strategy.

The rest of this paper is structured as follows. Sections 2 presents related and previous work to this study. Section 3 presents the research design and background information to the three case organizations. Section 4 presents the identified contribution objectives and complexities, and section 5 provides a discussion on the objectives and complexities in regards to related work. Section 6 presents a discussion on threats to validity, while section 7 concludes the paper. Appendices \ref{sec:appendix:secondCycle}--\ref{sec:appendix:complexities} provide detailed information about interview instruments and findings. 

\section{Related and Previous Work}\label{sec:rw}
This section provides an overview of related work to the two concepts of Contribution Objectives and Contribution Complexities. This is followed by an overview of contribution strategy research after which we present a summary of the related and previous work.

% Stuermer et al.~\cite{stuermer2009extending}
% * Giving up control

% Wnuk et al.~\cite{wnuk2012can}
% * Balance between contributing and reaping benefits
% * Unclear relationship between the benefits from contribution in terms of strategy and business goals

% West and Wood~\cite{west2008creating}
% * Managing conflicting needs of all stakeholders involved
% * Balancing the interests of those participants against those of the ecosystem leader

% Dahlander~\cite{dahlander2008firms}
% * challenge to aligning firm strategy with community

\subsection{Benefits of Sharing Software as OSS}\label{sec:rw:rationale}
% In their mapping of the potential benefits that may be gained for organizations using a commercial open source model~\cite{riehle2012single}, Sharivar et al.~\cite{sharivar2018business} divides the benefits into tangible revenues and intangible benefits. Tangible revenues are generated from sales in cases where an organization offers complementary products and services based on the OSS~\cite{sharivar2018business, andersen2012commercial, spijkerman2018open}, e.g., support and subscription offerings, proprietary extensions, or hardware-based products enabled by the OSS~\cite{chesbrough2007open, munir2017open, henkel2006selective}. Intangible benefits, however, are gained as an effect of sharing software OSS. Reasons for why an organization chooses to share software as OSS does, however, not have to be limited to one or the other~\cite{andersen2012commercial}.
Several studies have systematically surveyed the literature and to different extents covered the benefits for why software should be shared as OSS~\cite{sharivar2018business, munir2015open, host2011systematic, hauge2010adoption}. We categorize the benefits into four different themes.

A common theme is the cost-saving aspects~\cite{munir2018theory, munir2015open, andersen2012commercial}. By extending the resource-base~\cite{dahlander2008firms} and agreeing on a common standard~\cite{west2006challenges}, organizations can share the maintenance and quality assurance, accelerate the development and potentially decrease their time-to-release and market~\cite{munir2017open, stuermer2009extending, henkel2006selective, lindman2009beyond, olsson2017from, lundell2011practitioner}. By freeing up internal resources, they can focus on more value-adding activities~\cite{munir2017open, van2009commodification, lindman2009beyond}. On the opposite, by adopting a less symbiotic relationship to the OSS community~\cite{dahlander2005relationships}, an organization will have to maintain an internal branch of the OSS project, which may become costly depending on the number of modifications that need to be applied to new releases of the OSS project~\cite{wnuk2012can, ven2008challenges}. Hence, to attain these potential benefits, active engagement and a symbiotic relationship may be needed with the OSS community~\cite{chengalur2010empirical, dahlander2006man}.

Another common theme is the innovation aspects~\cite{munir2015open, andersen2012commercial}, which can be both product and process-oriented~\cite{munir2018theory}. By opening up the innovation process~\cite{chesbrough2007open} and ''pooling'' the R\&D/product development~\cite{west2006challenges}, organizations get access to an external workforce~\cite{lundell2010open}, which may bring increased knowledge sharing~\cite{nagle2018learning, lundell2011practitioner} and innovation at a lower cost~\cite{stuermer2009extending, ziegler2014why}. However, this external workforce should be seen as a complement rather than a substitute for internal knowledge and development~\cite{stam2009when, dahlander2008firms}. Munir et al.~\cite{munir2015open} describe it as a catalyst for ideas that may help organizations in broadening their offerings. Hence, an organization may question how much of its internal R\&D and innovation process it should outsource to a community~\cite{aagerfalk2008outsourcing}.

A third theme can be tied to improved reputation~\cite{munir2015open, ven2008challenges}. By creating a community or contributing to an existing community, an organization can create a marketing channel both towards (potential) customers, as well as future employees and have a positive effect on internal developers' satisfaction~\cite{lundell2010open, dahlander2008firms, riehle2011controlling, stuermer2009extending, henkel2006selective}. The improved reputation can turn into a competitive advantage~\cite{henkel2014emergence} and legitimize the use of the OSS from a public perspective~\cite{dahlander2006man}. An organization's customers are offered an opportunity to avoid vendor-lockin, and the ability to customize the software to internal needs~\cite{lundell2011practitioner, munir2017open}.

A fourth theme concerns control aspects~\cite{linaaker2018motivating}. If an OSS community has a meritocratic coordination process in place~\cite{shaikh2017governing}, influence on the development direction of the community may be gained by participating in the development and maintaining a symbiotic relationship~\cite{linaaker2019community, dahlander2005relationships, dahlander2006man, butler2018investigation, schaarschmidt2015firms, syeed2017measuring, nguyen2018do}. This may help steer the community including competitors and to manage potentially conflicting agendas~\cite{linaaker2019community, west2008creating, munir2015open, schaarschmidt2015firms, maenpaa2018organizing}.

Influence may also come implicitly when an organization's project or a feature is released and accepted as a standard solution, either within an existing community or as a new community~\cite{munir2017open}. If contributed to an existing community, other organizations will either have to accept and adapt, maintain internal forks of their solutions, or attempt to contribute their solutions in competition with the solution already established within a community~\cite{linaaker2018motivating}. If released as a new community and traction is gained, it can potentially become a new standard or compete with existing~\cite{lindman2009beyond, jansen2012shades, henkel2006selective, west2003open}, and create a surrounding ecosystem with complements from other organizations~\cite{west2007value, hartmann2016towards}.

Some of these themes may be more or less important depending on the type of organization. Munir et al.~\cite{munir2018theory} present a theory of openness that categorizes organizations using OSS in their tools and infrastructure setups based on why and when they adopt and share software as OSS. The ''why'' is either focused on reducing product development costs or building a symbiotic relationship with the OSS communities~\cite{dahlander2005relationships}. The ''when'' concerns whether a reactive or proactive strategy is adopted. In the former, an organization adapts to existing and upcoming OSS communities without taking any initiatives. In the latter, an organization has a long-term agenda and adapts its OSS community engagements or create new communities accordingly.

\subsection{Costs and Risks of Sharing Software as OSS}
Deciding what should be contributed is a complex matter~\cite{munir2015open}. Even though the amount of a software that may be considered differentiating is often limited~\cite{lindman2009beyond, van2009commodification}, the risk of sharing differentiating functionality and sensitive Intellectual Property Rights (IPR), and as a consequence losing a competitive edge, is a recognized challenge in literature (e.g.,~\cite{wnuk2012can,van2009commodification, henkel2006selective,henkel2008champions,west2006challenges,iivari2008usability}). Instead of disqualifying a complete software artifact however, one approach may be to selectively reveal commodity or enabling parts while keeping differentiating parts closed~\cite{west2003open, henkel2006selective, stuermer2009extending}. An alternative approach may be to ''spinout''~\cite{west2006challenges} disclose the software artifact under a restrictive copy-left license~\cite{west2003open}, e.g., the General Public Licence version 2 and 3, or the Affero GPL\footnote{https://opensource.org/licenses/alphabetical}. This is a common approach for commercial OSS organizations~\cite{riehle2012single} using a dual-license approach to appropriate and capture value from customers~\cite{chesbrough2007open}. Combinations can be found, e.g., where a core OSS project is permissively licensed, while certain extensions or improvements are licensed with more restrictive licenses~\cite{deodhar2012strategies}, or kept proprietary~\cite{weikert2013model}.

A related challenge is determining when software should be contributed~\cite{wnuk2012can}. Several studies have attempted to model and identify an optimal timing~\cite{haruvy2008open, kort2011should, caulkins2013when}. Caulkins et al.~\cite{caulkins2013when} for example, identify costs related to the development and adapting the business model, along with the software quality as factors affecting when software should be released as OSS. Quality is in this case not just referred to the number of errors, but to the software's features and functionality. The shift in how quality changes can be compared to where in the commoditization process a software artifact is. I.e., the \textit{``software artifact's value depreciation and how it moves between a differential to a commodity state, i.e., to what extent the artifact is considered to help distinguish the focal organization's product offering relative to its competitors''}~\cite{linaaker2018motivating}. As suggested by van der Linden et al.~\cite{van2009commodification}, a first step may be to open up the software in joint ventures or closed strategic alliances~\cite{linaaker2018motivating}, while a third step may be to release it as OSS. Wnuk et al.~\cite{wnuk2012can} describe the challenge as a balance between losing a competitive edge and increased maintenance costs.

Where to contribute is a third challenge. When contributing the software artifact to an existing OSS community not governed by the organization itself~\cite{o2007governance}, the organization needs to consider the requirements engineering process of the community~\cite{scacchi2002understanding}. In this context, the organization is a stakeholder among many and needs to consider potentially conflicting agendas from other stakeholders~\cite{munir2015open, schaarschmidt2015firms, maenpaa2018organizing, linaaker2019method}. If there is a misalignment between the organization's and the community's Requirements Engineering (RE) process~\cite{ernst2012case, alspaugh2013ongoing, kuriakose2015how}, the organization may need to influence the development direction of the community~\cite{munir2017open}. If the organization lacks the influence needed, they need to consider the cost of gaining it~\cite{linaaker2019community}. An option is to release the software as an independent OSS project and build a new community around it, which may require significant investment as well~\cite{kilamo2012proprietary, west2005contrasting, dahlander2008firms}.

\subsection{Contribution Strategies}\label{sec:rw:CAP}
Wnut et al.~\cite{wnuk2012can} highlighted the importance of contribution strategies early on. Research on the topic has however been limited~\cite{munir2015open} with some exceptions~\cite{wnuk2012can, weikert2013model, linaaker2018motivating}.

\begin{figure}
\begin{center}
\includegraphics[width=\columnwidth]{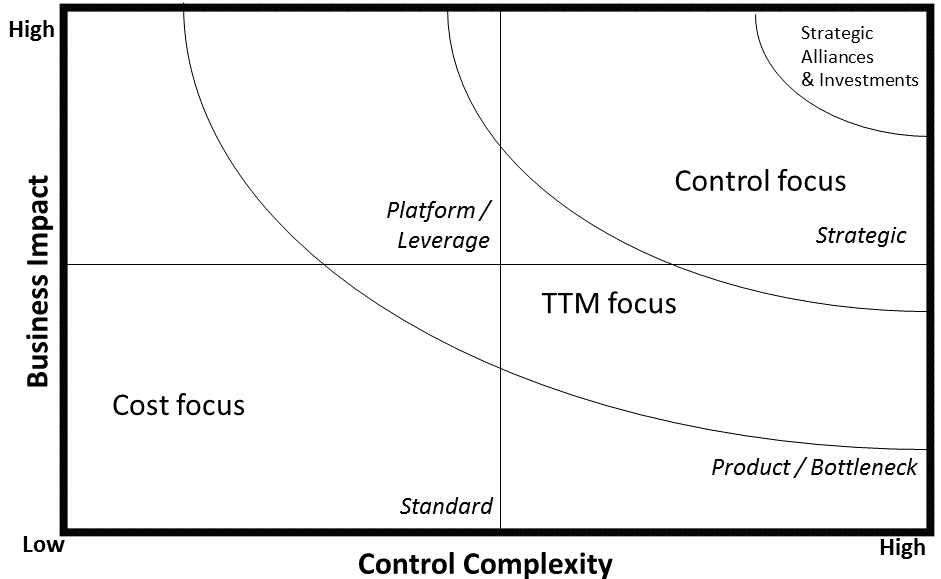}
\label{fig:OImodel}
\caption{The Contribution Acceptance Process (CAP) model as presented in Lin{\aa}ker et al.~\cite{linaaker2018motivating}. Based on how a software artifact is valued in regards to its business impact and control complexity, a contribution strategy is elicited pending the artifact's placement on the grid.}
\end{center}
\end{figure}

In previous work~\cite{linaaker2018motivating}, the first author of this study conducted a case study on the contribution strategy decision process at Sony Mobile. Through the case study, a Contribution Acceptance Process (CAP) model was designed with the purpose to help organizations decide if a software artifact should be shared as OSS. A software artifact (e.g., features or projects) is valued according to its \textit{business impact} (how much you profit from the component) and \textit{control complexity} (how hard the technology and knowledge behind the artifact is to acquire and control). With the help of a series of questions provided, the software artifact could be ranked qualitatively and placed in a two-by-two matrix (see Fig.~\ref{fig:OImodel}), with each cell representing a certain type of generic artifact with its own contribution strategy that is proposed for the artifact.

As an example, one of the four artifact types concerned strategic artifacts (upper right quadrant in Fig.~\ref{fig:OImodel})~\cite{linaaker2018motivating}. These artifacts have a high business impact and a high control complexity. They contain differentiating value and provides a competitive edge to the organization. Differentiating parts should be kept closed while enabling parts, should be contributed. The contribution is recommended to be made to a community where the organization has a high level of influence. If not possible, a new community may be created.

The contribution strategy could then be fine-tuned based on a series of objectives~\cite{linaaker2018motivating}. The more strategic an artifact is (higher business impact and control complexity) the faster the artifact (or enabling parts of it) should be contributed to establishing the artifact as the standard solution. The organization could thereby avoid having to adapt to competing solutions and instead strive towards steering its competitors. For this reason, these kinds of artifacts should be prioritized before standard artifacts, which may be considered as standard knowledge.

An example of a strategic artifact included a multimedia framework that enabled differentiating camera functionality~\cite{linaaker2018motivating}. The framework was contributed, while the camera functionality could be kept closed. This gave Sony Mobile the advantage of not having to refactor or adapt to competing frameworks and still keep differentiating functionality closed.

The CAP model can be used both in a proactive and reactive approach~\cite{linaaker2018motivating}. In the proactive, it is used to map a series of software artifacts, e.g., features in the product planning process~\cite{kittlaus2017software}. A cross-functional team may be assembled, including e.g., representatives from marketing, legal and product development. The team can iterate the mapping process comparing features against each other through consensus-seeking discussions. In the reactive approach, the CAP model is used in a similar manner but for making decisions in regard to incoming contribution requests from the organization's development teams.

The validation of the CAP model showed that the CAP model provided a good foundation for discussion. Feedback pointed out that the questions and scale used to value a software artifact in terms of business impact and control complexity, was found useful but in need of being tailored to the context where the CAP model is applied. A highlighted concern for future work was to avoid making the following versions of the CAP model more complex.

% \subsection{Contribution Processes}\label{sec:rw:process}
A contribution process starts with an individual filing a contribution request, requesting to be allowed to contribute a certain software artifact to a certain community or create a new community. The request is then managed by an entity within the organization that has the mandate to decide on a contribution strategy.

In the case of Sony Mobile~\cite{munir2017open, linaaker2018motivating, mols2017open}, an individual from the development organization fills in a contribution request answering a set of pre-defined questions. One of these questions concerns the size and complexity of the contribution~\cite{linaaker2018motivating}.

\begin{itemize}
    \item \textit{Trivial contributions} include small changes such as bug fixes and minor improvements to existing OSS communities.
    \item \textit{Medium contributions} include new features and larger architectural changes to existing communities.
    \item \textit{Major contributions} include projects where a new community is to be established or software artifacts containing important IPR such as patents.
\end{itemize}

Trivial contribution requests are decided on by the closest manager, while medium and major contribution requests are processed by an Open Source Governance board~\cite{linaaker2018motivating}. The board has a cross-functional constellation with competencies covering aspects such as legal, user experience, product development, and product ownership~\cite{munir2017open}. Each medium and major contribution request is investigated further, which includes an IPR review rendering in a decision from the board on a relevant contribution strategy~\cite{linaaker2018motivating}. To ease bureaucracy, the board can develop general contribution strategies for specific communities allowing developers to, e.g., contribute minor and medium contributions to an OSS community without having to submit a request. This approach was applied to OSS communities where the OSS project was considered not providing a competitive edge to Sony Mobile, e.g, the two development-tools Jenkins and Gerrit~\cite{munir2017open}.

Creating awareness and maintaining the contribution process can be a challenge in large organizations~\cite{wnuk2012can}. In Sony Mobile, the contribution process and Open Source Governance board are led and supervised by the Open Source Program Officer, a role that connects management, legal, IPR and software development functions around Sony Mobile's OSS operations~\cite{mols2017open}. Similar organizational and process setups have been reported on in the documents and guidelines published by the TODO-group\footnote{https://todogroup.org/}, a foundation for organizations that has an established Open Source Programs Office. The program's office can be viewed as an organizational entity including the Open Source Program Officer and any further roles tied to an organization's OSS operations, e.g., concerning compliance and community management. In his more compliance-focused overview of OSS governance within an organization, Kemp~\cite{kemp2010open} proposes the creation of an ''OSS working party'' and an ''OSS compliance officer'', which to some extent can be compared to an Open Source Programs Office and an Open Source Program Officer.

\subsection{Summary}
The benefits that may incentivize an organization to contribute its software as OSS are many~\cite{sharivar2018business, munir2015open, host2011systematic, hauge2010adoption}, as are the potential costs and risks that may remove or outweigh the benefits~\cite{munir2015open, linaaker2015requirements}. To gain the expected benefits, an organization, therefore, needs to consider the contribution objectives and complexities relevant to their context, and make an informed decision on what they contribute, where, and when, in alignment with their business goals. Related work on how organizations can develop such strategies is limited~\cite{munir2015open}, with some exceptions~\cite{wnuk2012can, weikert2013model}, including our previous work with Sony Mobile and the CAP-model~\cite{linaaker2018motivating}. The validation of the CAP-model pointed to a need for more general and less complex approaches for creating contribution strategies. This study, therefore, aims to identify and define a set of contribution objectives and complexities from which organizations can choose those relevant and thereby create contribution strategies based on their specific contexts and business goals.

\section{Research Design}\label{sec:rd}
To answer the research questions \textbf{RQ1} and \textbf{RQ2} as defined in section~\ref{sec:introduction}, we employed a multiple-case study~\cite{runeson2012casestudy} in an iterative approach with two research cycles. The method offers a way for generating in-depth knowledge and understanding of a phenomena in how and why it occurs~\cite{easterbrook2008selecting}. In our study, this regards an exploratory investigation of what Contribution Objectives (COs) and Contribution Complexities (CCs) that the case organizations consider in decisions on whether a software artifact should be released as OSS, when in time, and if it should be contributed to an existing community or if a new community should be created. Answers to these questions together form the contribution strategy for the software artifact. The decision of arriving at such strategies makes up our unit of analysis~\cite{runeson2012casestudy}.

In total three case organizations were studied, one in the first research cycle and two in the second. Through this approach, we could develop the first set of contribution objectives and complexities which could then be used as a foundation when entering the second research cycle. Below we first describe the case organizations. We then describe how the research was carried out through the two research cycles.

\subsection{Case Organizations}\label{sec:caseOrgs}
Below we describe and provide context to each of the three case organizations studied, denoted CaseOrg1-3. Per organization, we present a general description, along with a more in-depth overview of their contribution process as well as examples of OSS projects, which they have released, or are active contributors to.

\subsubsection{CaseOrg1}
\textbf{General description:} CaseOrg1 is a US-based media and technology company providing video, high-speed internet, smart home and voice services. They have 1000+ employees and develop their own software to enable and deliver their services to the customers. Having been passive consumers of OSS before 2006, they became active contributors starting in 2006. Since then, they have released several OSS projects and are active contributors in several others, as well as members of a number of OSS foundations.

\textbf{Contribution process:} Internally, they have an Open Source Programs Office set up to develop and manage e.g., contribution and compliance processes, and community management. Their contribution process starts with a developer filling out a contribution request form concerning a software artifact (e.g., feature or project). If the artifact regards a smaller contribution (e.g., a bug fix, or documentation), or a contribution to an OSS community deemed generally as non-competitive, approval can be gained online by the manager and the Open Source program office. For more significant components, the contribution request is managed by an OSS advisory board, which has representatives from legal, IPR, development and business functions within the organization. The board then gives a final decision on how to proceed. CaseOrg2 also has what they refer to as \textit{sandbox approval}, which allows a project and identified contributors to be approved to contribute to a project without coming back for each patch. This governance set-up and contribution process is similar to that of Sony Mobile~\cite{linaaker2018motivating}.

\textbf{Example OSS projects:} One example concerns an internally developed Linux distribution that is embedded in hardware devices shipped to customers, which enables the delivery of CaseOrg1's consumer-oriented services. The software was initially developed to replace a proprietary option and was later released as OSS under the governance of an industry consortium. Another example concerns an infrastructure project, which enables the delivery of services related to secure and reliant internet-traffic to business-oriented customers. The project was released under the governance of a neutral community with an existing OSS foundation. A third example regards a DNS-as-a-Service project originally developed to increase internal operational efficiency. The decision process is thoroughly reported on in previous work~\cite{linaaker2018motivating}.

\subsubsection{CaseOrg2}

\textbf{General description:} CaseOrg2 is a European-based hardware electronics manufacturer serving both business and private customers. They have 1000+ employees and develop their own software and orchestrate a software ecosystem~\cite{jansen2009sense} to enable and deliver their services to the customers. This study has focused on its Tools department which develops and maintains development tools and infrastructure projects used by the organization's product development teams. A majority of the tools and infrastructure projects are OSS-based with active engagement from the Tools department in their respective communities. The active engagement includes continuous contributions of features and plugins to existing OSS communities as well as the release of new OSS projects.

\textbf{Contribution process:} CaseOrg2 does not have a dedicated Open Source Programs Office. The organization has two contribution processes set up, one for their product development teams, and one for the Tools department. The latter is less strict than the former as CaseOrg2 generally considers the projects developed within the Tools department to provide a limited competitive edge to the organization. The contribution process used within the Tools department is initialized by a developer filling out a contribution request form concerning a software artifact (e.g., feature or project). If the contribution request is intended for an existing community, the request is managed by the department manager. If the contribution request concerns the creation of a new OSS community, the request is managed by the business unit manager. If deemed necessary the concerned manager will consult relevant functions within the organization such as Legal and IPR departments.

\textbf{Example OSS projects:} The Tools department has contributed and maintains several plugins to the Jenkins\footnote{https://jenkins.io/} and Gerrit\footnote{https://www.gerritcodereview.com/} OSS projects. Jenkins is a build-server and Gerrit is a code-review tool, both used in CaseOrg2's continuous integration tool-chain. The Tools department has recently also started to create new OSS projects that fall outside existing communities. One such project was developed internally to allow for the creation and use of shorter URLs to internal resources. These are maintained as standalone projects on CaseOrg2's Github\footnote{https://github.com/} page.

\subsubsection{CaseOrg3}

\textbf{General description:} The Swedish Public Employment Service\footnote{https://arbetsformedlingen.se/} makes up the third case organization. They are non-anonymous but are referenced to as CaseOrg3 for consistency. CaseOrg3 is a public sector agency in Sweden with the main goal to facilitate and enable matching between job-seekers and employers on the Swedish labor market. The organization has 10 000+ employees of which 600 are employed within their IT division. The focus of this study is on a department within the IT division which aims to create a platform\footnote{https://jobtechdev.se/} on which private actors can build complementary products and services for matching job-seekers and employers. The platform, consisting of OSS projects and Open Data sources, is intended to help CaseOrg3 to move from the role of being a service provider to become a service enabler and help the platform's ecosystem members to collaborate and co-create, potentially resulting in accelerated innovation and a more efficient job-matching on the Swedish labor market.

\textbf{Contribution process:} CaseOrg3 does not have a dedicated Open Source Programs Office or contribution process in place. The studied department prioritize the release of internal projects based on what they believe is of most value to the platform's ecosystem of private actors.

\textbf{Example OSS projects:} The studied department has released a number of OSS projects consisting of small components that can integrate into existing applications, stand-alone end-user applications, and developer-focused tools and utilities\footnote{https://jobtechdev.se/doc/samples/}. One project is a video-conference-tool used internally at CaseOrg3 to facilitate remote interviews for employers and job-seekers. Another example concerns a search engine that can be used to access and search among job listings in CaseOrg3's Open Data sources with job ads.

\subsection{Research Cycle 1}\label{sec:rd:firstCycle}
In the first research cycle, we investigated the problem context through a first case study of CaseOrg1. Data was gathered through six semi-structured interviews with employees from different areas of the organization, see Table~\ref{tbl:interviewees}. A questionnaire (see Appendix~\ref{sec:appendix:secondCycle}) was created based on findings from an earlier reported case study of the contribution strategy decision process at Sony Mobile~\cite{linaaker2018motivating} conducted by the authors of this study.

\begin{table}[b!]
\caption{List of interviewees from CaseOrg1-3.}
\label{tbl:interviewees}
\begin{tabular}{@{}llll@{}}
\toprule
\textbf{ID} & \textbf{Case organization} & \textbf{Title} & \textbf{Employment} \\ \midrule
I1 & CaseOrg1 & Open Source Program Officer & 3 years \\
I2 & CaseOrg1 & Community Manager & 9 years \\
I3 & CaseOrg1 & Director of Software Development & 9 years \\
I4 & CaseOrg1 & Director of Software Architecture & 14 years \\
I5 & CaseOrg1 & Vice President - Standards & 16 years \\
I6 & CaseOrg1 & Chief Software Architect & 13 years \\
I7 & CaseOrg2 & Project Manager & 1 year \\
I8 & CaseOrg2 & Senior Developer & 4 years \\
I9 & CaseOrg2 & Junior Developer & 1 year \\
I10 & CaseOrg2 & Department Manager & 5 years \\
I11 & CaseOrg2 & Business Unit Manager & 1 year \\
I12 & CaseOrg3 & External Consultant & 3 years \\
I13 & CaseOrg3 & Chief Digital Officer & 6 years \\
I14 & CaseOrg3 & Department Manager & 6 years \\
I15 & CaseOrg3 & Product Owner & 9 years \\
I16 & CaseOrg3 & Project Manager & 2 years \\
I17 & CaseOrg3 & Community Manager & 1 year \\
I18 & CaseOrg3 & Security Engineer & 1 year \\
I19 & CaseOrg3 & Senior Developer & 6 years \\
I20 & CaseOrg3 & Senior Developer & 5 years \\ \bottomrule
\end{tabular}
\end{table}

Interviewees were selected in order to gain different and complementary perspectives on CaseOrg1's contribution strategy decision process. Each of the interviews was audio-recorded and transcribed. An anonymized interview summary was presented to I1 and I2 to verify interpretations and clarify any misunderstanding. The transcripts were then coded with an inductive approach and audit trails maintained~\cite{runeson2012casestudy}. Using an iterative and refining coding process~\cite{runeson2012casestudy, robson2011real}, sentences and paragraphs from the interviews were given descriptive topic sentences which then merged together in common codes and sorted under the two categories contribution objectives and complexities, mapping to \textbf{RQ1} and \textbf{RQ2} respectively. This rendered in a first set with a total of 16 objectives and 12 complexities (see table~\ref{tbl:coding}).

Below we provide an example coding from CaseOrg1, which later was conceptualized as Contribution Objective CO3 after the updated coding from the second research cycle:
\begin{itemize}
\item \textbf{Code}: Developer satisfaction
    \begin{itemize}
    \item \textbf{Quote by I5}: \textit{``\ldots  I think a real reason goes to staffing engagements and retention. It is important to people and I think a positive employment characterization that you get to engage with open source communities and that the company does release something as open source projects. I think that's a big selling point that people are looking for in whom they want to work for as an engineer.''}.
    \end{itemize}
\end{itemize}

\subsection{Research Cycle 2}\label{sec:rd:secondCycle}
In the second research cycle, our goal was to validate the contribution objectives and complexities identified in the previous cycle while continuing to explore the problem context and identify new ones. The interview questionnaire was therefore updated based on previous findings. To validate the questionnaire and gain further input into its revision, we studied \textit{contribution request forms} collected from seven different software-intensive organizations (see Table~\ref{tbl:organizations}). A contribution request form contains questions about software artifacts, and often guidelines for, and examples of what types of contributions the organization generally accepts. The form is often submitted by the engineer or team behind the concerned software artifact. The contribution request forms give an indication of what the organizations consider when deciding whether the software artifact should be contributed.

\begin{table}[htbp]
\caption{List of organizations from where contribution request forms has been gathered.}
\label{tbl:organizations}
\begin{tabular}{@{}lllll@{}}
\toprule
\textbf{ID} & \textbf{Business} & \textbf{Market} & \textbf{Employees} & \textbf{Use of OSS}\\ \midrule
O1 & Consumer electronics & Worldwide & 4 000+ & Infrastructure \& Products\\
O2 & Consumer electronics & Worldwide & 100 000+ & Infrastructure \& Products \\
O3 & Software products & Worldwide & 30 000+ & Infrastructure \& Products \\
O4 & Software products & Worldwide & 100 000+ & Infrastructure \& Products \\
O5 & Telecom & North America & 1 000+ & Infrastructure \& Products \\
O6 & Consumer electronics & Worldwide & 1 000+ & Infrastructure \& Products  \\
O7 & Industry organization & North America & - & - \\ \bottomrule
\end{tabular}
\end{table}

The revised questionnaire (see Appendix~\ref{sec:appendix:thirdCycle}) was then used in semi-structured interviews with five employees from CaseOrg2 and nine employees from CaseOrg3. As in the former cycle, interviewees were selected in order to gain different and complementary perspectives on two organizations' contribution strategy decision process. All interviews were audio-recorded and transcribed. Anonymized interview summaries were presented to I7 and I8 from CaseOrg2 and I15 from CaseOrg3.

Using the coding schema from the first cycle, transcripts from CaseOrg2 were first coded, which resulted in three new COs and one CC being added. In the following step, the new coding schema covering CaseOrg1-2 was applied to transcripts from CaseOrg3, resulting in one new CO being added, six COs merged into three. Three new CCs were added, of which two were former COs. Transcripts from all case organizations were then reiterated, and a final coding scheme assembled. This resulted in two COs being merged into one, two COs converted to CCs, and six CCs being merged into three. In table~\ref{tbl:coding}, the evolution, and saturation of the COs and CCs throughout the coding process are presented.

\begin{table}[htbp]
\caption{Evolution and saturation of COs and CCs throughout the coding process consisting of four steps. The first coding schema was based on transcripts from CaseOrg1. This was then applied and revised based on transcripts from CaseOrg2, and then CaseOrg3. The coding schema was then reiterated, resulting in the final set of COs and CCs.}
\label{tbl:coding}
\begin{tabular}{@{}lllll@{}}
\toprule
\textbf{} & \textbf{1. CaseOrg1} & \textbf{2. CaseOrg2} & \textbf{3. CaseOrg3} & \textbf{4. Final coding}\\ \midrule
COs (New)       & 16    & 3     & 1     & 0 \\
COs (Merged)    & 0     & 0     & 6 \textrightarrow 3     & 2 \textrightarrow 1 \\
COs (Removed)   & 0     & 0     & 2 \textrightarrow CC     & 2 \textrightarrow CC \\\midrule
COs (Total)     & 16    & 19    & 15    & 12 \\\midrule
CCs (New)       & 12    & 1     & 3     & 2 \\
CCs (Merged)    & 0     & 0     & 0     & 6 \textrightarrow 3 \\
CCs (Removed)   & 0     & 0     & 0     & 0 \\\midrule
CCs (Total)     & 12    & 13    & 16    & 15 \\ \bottomrule
\end{tabular}
\end{table}

Below we present a continuation of the example code \textit{Developer satisfaction} provided in Section~\ref{sec:rd:firstCycle}. After coding the transcripts from CaseOrg2-3, further support was identified for \textit{Developer satisfaction}. This code was then consolidated with the new code \textit{Hire talent}, which was also identified in transcripts from CaseOrg1 in the reiteration, finally rendering in contribution objective CO3.

\begin{itemize}
    \item \textbf{Contribution Objective CO3}: Improve employer branding 
        \begin{itemize}
        \item \textbf{Code}: Developer satisfaction
            \begin{itemize}
            \item \textbf{Quote by I5}: \textit{``\ldots  I think a real reason goes to staffing engagements and retention. It is important to people and I think a positive employment characterization that you get to engage with open source communities and that the company does release something as open source projects. I think that's a big selling point that people are looking for in whom they want to work for as an engineer.''}.
            \item \textbf{Quote by I19}: \textit{``We think it is fun and it is positive for our personal satisfaction to contribute back''}.
            \end{itemize}
        \item \textbf{Code}: Hire talent
            \begin{itemize}
            \item \textbf{Quote by I8}: \textit{``And then as a result of that, we already see it in some areas, because of our involvement in open standards or open source communities, we're able to attract a different type of engineer, or a different higher level engineer or value-add that they can bring to the company as a result of our openness and engagement. And so it's sort of a natural reinforcing cycle.''}.
          \item \textbf{Quote by I7}: \textit{``Show that we are a modern firm with a presence in open source and that we contribute and not just consume, which attracts a lot of great developers and becomes a channel for us to attract new employees.''}.
            \end{itemize}
      \end{itemize}
\end{itemize}

The revised set of COs and CCs were presented and discussed with I1 from CaseOrg1, I7 and I8 from CaseOrg2 and I15 from CaseOrg3. All interviews were audio-recorded and transcribed. The interviewees were first asked about the correctness of the COs and CCs that had been mapped to their organization. Interviewees were then asked if any COs or CCs not mapped to them were found relevant for the organization. Lastly, the interviewees were asked about the general completeness, applicability, and usability of the set of COs and CCs, and whether something was redundant, missing, or could potentially be modified. Existing COs and CCs were then refined based on interview findings, while none were added nor removed.

\section{Results}\label{sec:results}

% Compare to Ibrahim's classification
% http://www.ibrahimatlinux.com/uploads/6/3/9/7/6397792/oss-strategy-objectives.pptx

\begin{figure}[htbp]
\centering
\includegraphics[width=1\textwidth]{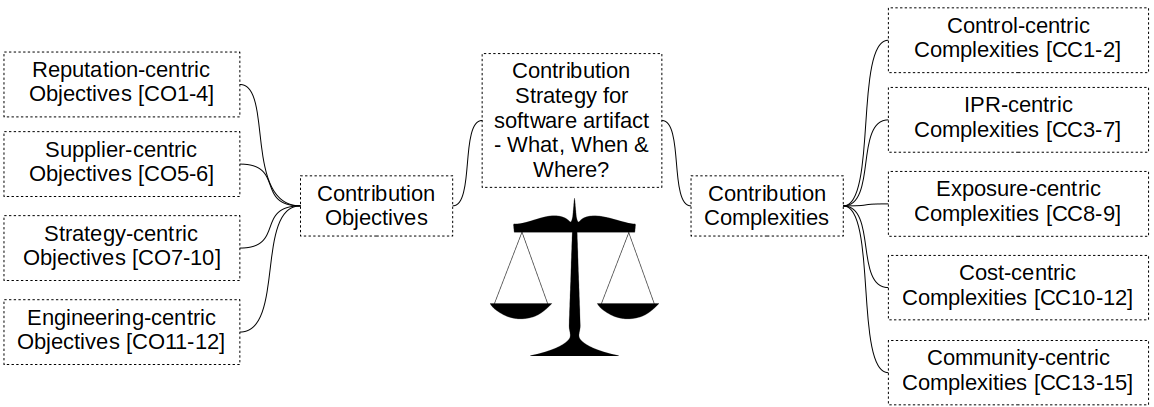}
\caption{This study presents 27 considerations (12 Contribution Objectives (CO) and 15 Contribution Complexities (CC)) that may  need to be considered by an organization when deciding on a contribution strategy for a software artifact. The COs and CCs are divided into four and five categories respectively and are listed in Tables~\ref{tbl:contributionObjectives} and~\ref{tbl:contributionComplexities}.
}
\label{fig:CO-CC-map}
\end{figure}

In this section, we summarize the identified contribution considerations, grouped into Contribution Objectives (COs) and Contribution Complexities (CCs), which an organization may analyze and weigh against each other when deciding on a contribution strategy for a certain software artifact. In total, we present 12 COs and 15 CCs divided over four and five categories respectively (see Fig.~\ref{fig:CO-CC-map}).

% We define a \textbf{contribution objective} as \textit{a purpose for contributing a software artifact, motivated by a monetary or non-monetary benefit that is enabled or resulted directly or indirectly as a consequence of the contribution}.

% A \textbf{contribution complexity} we define as \textit{an aspect related to a software artifact that may complicate the contribution of the artifact, or imply a cost or risk as a result thereof, either directly or indirectly}.\Note{Borde dessa definitioner komma i samband med forskningsfrågan i stället så man inte tror de är empiriska utan fattar att de är utgångspunkt?}
% The complexities are further divided into two categories: software artifact-centric and community-centric contribution complexities. The former category focuses on aspects related to the software artifact under investigation, while the latter focuses on aspects related to the intended recipient of the software artifact, whether it be an existing or a new OSS community.

All contribution considerations are maybe not relevant to all organizations and all software artifacts. To help guide decision-makers and those making contribution requests, the presented contribution considerations can help as input when forming an organization-specific contribution strategy. COs and CCs can then be described in the context of the focal organization and relevant questions asked in a contribution request form when setting up a contribution process within the organization. An individual intending to file a request is then enabled to beforehand understand the rationale used by the decision-makers when deciding on a contribution strategy, and thereby to provide motivated arguments why the request should be accepted. For further discussion, see Section~\ref{sec:disc}.

The COs are presented, each with a number of (potential) key benefits, in Table~\ref{tbl:contributionObjectives}, while the CCs are presented, each with a number of key concerns, in Table~\ref{tbl:contributionComplexities} and are further explained below. A detailed description of each consideration, together with example quotes from interviews, are given in Appendices~\ref{sec:appendix:objectives} and~\ref{sec:appendix:complexities}. 

% The contribution objectives of the CMF are presented, each with a number of (potential) key results, in Table~\ref{tbl:contributionObjectives}.
% The software artifact-centric and community-centric contribution complexities are presented in Table~\ref{tbl:softwareCentricContributionComplexities} and Table~\ref{tbl:communityCentricContributionComplexities} respectively, each presented with a number of key concerns.

\subsection{Contribution Objectives}
In total, 12 COs were identified in the analysis process divided over four categories: reputation-centric objectives, supplier-centric objectives, strategy-centric objectives, and engineering-centric objectives. Below an overview is given to the COs in each category. A more detailed overview is presented in Appendix~\ref{sec:appendix:objectives} where the COs are described separately with examples per identified benefit. Quotes from interviewees of the case organizations are given in Appendix~\ref{sec:appendix:objectives} to provide context for each of the examples.

\begin{table*}[htbp]
\caption{Overview of the contribution objectives, related key benefits, in which case organization they were identified, and examples of similar findings in literature.}
\label{tbl:contributionObjectives}
\begin{tabular}{p{0.4cm} p{1.7cm} p{6.2cm} p{0.5cm} p{1cm}}
\toprule
\textbf{ID} & \textbf{Contribution \newline objective} & \textbf{Key benefits} & \textbf{Case \newline Org.} & \textbf{Lit.\newline ref.} \\ \midrule

\multicolumn{5}{@{} l}{\hspace{0.2cm}\textbf{Reputation-centric Objectives}} \\ \midrule

CO1 &
Prove skill and influence &
$\bullet$ Improved trust towards customers. \newline
$\bullet$ Improved trust towards partners. & 1,2 &
~\cite{henkel2006selective, stuermer2009extending, dahlander2008firms, linaaker2019community} \\ \midrule

CO2 &
Increase transparency &
$\bullet$ Improved trust among customers. (CaseOrg1) \newline
$\bullet$ Improved trust among the public. (CaseOrg3) & 1,3 &
~\cite{dahlander2006man, stuermer2009extending, linaaker2019community} \\ \midrule

CO3 &
Improve employer branding &
$\bullet$ Attraction of talented employees. \newline
$\bullet$ Retention of existing employees. & 1,2,3 &
~\cite{henkel2006selective, stuermer2009extending, linaaker2019community}\\ \midrule

CO4 &
Be a good open source citizen &
$\bullet$ Idealistic satisfaction among employees. & 1,2,3 &
~\cite{henkel2006selective, jansen2012shades, linaaker2019community} \\ \midrule

\multicolumn{5}{@{} l}{\hspace{0.2cm}\textbf{Supplier-centric Objectives}} \\ \midrule

CO5 &
Create price pressure &
$\bullet$ Lower subscription costs of procured products and services. \newline
$\bullet$ Lower prices on tenders. & 1,3 & \\ \midrule

CO6 &
Outsource infrastructure operation &
$\bullet$ Internal focus on activities related to core business. & 1 & \\ \midrule

\multicolumn{5}{@{} l}{\hspace{0.2cm}\textbf{Strategy-centric Objectives}} \\ \midrule

CO7 &
Collect data &
$\bullet$ Improved machine learning and artificial intelligence algorithms. \newline
$\bullet$ Creation of solutions based on Open Data sources. & 1,3 & \\ \midrule

CO8 &
Standardize solution &
$\bullet$ Force competitors to adapt and steer market or community according to internal agenda. (CaseOrg1-2)\newline
$\bullet$ Improve competition and enable more value-adding development on market. (CaseOrg3) & 1,2,3 &
~\cite{lindman2009beyond, jansen2012shades, henkel2006selective, west2003open, linaaker2019community, linaaker2018motivating} \\ \midrule

CO9 &
Build a software ecosystem &
$\bullet$ Enable and stimulate creation third party applications and services. & 2,3 &
~\cite{west2007value, west2008creating, hartmann2016towards} \\ \midrule

CO10 &
Improve partner collaboration &
$\bullet$ Increased value of software ecosystem. & 2 & \\ \midrule

\multicolumn{5}{@{} l}{\hspace{0.2cm}\textbf{Engineering Objectives}} \\ \midrule

CO11 &
Open up innovation process &
$\bullet$ Accelerated innovation process. (CaseOrg1-3) \newline
$\bullet$ Creation of more and better market-oriented solutions. (CaseOrg3) & 1,2,3 &
~\cite{munir2015open, andersen2012commercial, linaaker2018motivating, linaaker2019community}\\ \midrule

CO12 &
Extend development resources &
$\bullet$ Focus on more value-adding development. \newline
$\bullet$ Accelerated development. \newline
$\bullet$ Lower maintenance cost. & 1,2,3 &
~\cite{munir2017open, linaaker2018motivating, stuermer2009extending, henkel2006selective, lindman2009beyond, olsson2017from}\\ \bottomrule

\end{tabular}
\end{table*}

\subsubsection{Reputation-centric Objectives}
The reputation-centric objectives highlight the value in creating and maintaining a good reputation towards different stakeholders, including customers, partners, community, as well as existing and potential employees. By contributing to a specific community, an organization can improve the trust of customers and partners. The organization can demonstrate relevant technical knowledge about the OSS, and gain the influence needed in order to steer the development in a way that aligns with expectations and wishes from the customers and partners. Both retention and attraction of new customers and partners are further mentioned as potential results (CO1).

Increasing transparency is an aligning objective (CO2) in order to build trust among customers, or even the public. Being transparent about how the quality of a service is measured or what is produced based on public funding are two examples that are highlighted in interviews.

Allowing and enabling engineers to contribute to OSS can benefit the organization both in terms of retention and attraction of new talent (CO3). Reported examples include the ability to build a public CV, having an impact and collaborating with people outside the organization. Idealistic satisfaction of ''doing the right thing'' by contributing back and not just consuming OSS is also viewed as an important aspect among an organization's employees (CO4).

\subsubsection{Supplier-centric}
Supplier-centric objectives focus on how an OSS software artifact can be leveraged to better extract value from supplier-relations. For example, releasing projects as OSS can be a way to put price pressure on those providing corresponding products, as well as a means of inviting more potential suppliers to bid on tenders and thereby make the price more competitive (CO5). However, price pressure does not have to be the main incentive in terms of suppliers. By making a project available as OSS it becomes possible for cloud providers to run and maintain the project at scale and thereby lower the cost, while enabling internal resources to focus on more value-adding activities (CO6).

\subsubsection{Strategy-centric Objectives}
The strategy-centric objectives consider contributions that can have a larger impact on the organization's ability to stay competitive in the business environment where it operates~\cite{dasilva2014business}. Releasing a software artifact as OSS may, for example, enable generation and collection of data that can be used to improve machine learning and artificial intelligence algorithms, while also enabling the potential development of new products or services based on the data (CO7).

A more control-focused objective (CO8) concerns the creation of a standard solution, either within an industry or within a specific community. If a solution gains traction and acceptance, it could provide a first-mover advantage as other actors would have to adapt. It may also provide the organization with some level of influence on the development of the artifact, with the potential to steer the direction of either a community or industry.

A related and potentially overlapping objective (CO9) may be to create a software ecosystem~\cite{jansen2009sense} with the rationale of attracting and enabling third-party developers to create complementary products and services, and potentially also contribute to the platform constituted by the OSS project released by the platform provider. For example, releasing tools and infrastructure projects related to the platform, could further help to improve collaboration with partners and third-party developers, and thereby help increase the value of the software ecosystem (CO10).

\subsubsection{Engineering-centric Objectives}
The engineering-centric objectives concerns contributions where the rationale is to explicitly exploit the knowledge and resources of a community to one's advantage. One such objective (CO11) is to open up the innovation process by using the concerned community as a source of e.g., new and innovative requirements and feature implementations. An overlapping objective (CO12) is to use the development resources of the community as a force multiplier in order to share maintenance and quality assurance, while also accelerating development and enabling a value-adding focus for engineers internally of the organization.

\subsection{Contribution Complexities}
In total, 15 CCs were identified in the analysis process divided over five categories: control-centric complexities, IPR-centric complexities, exposure-centric complexities, cost-centric complexities, and community-centric complexities. Below, an overview is given to the CCs in each category. A more detailed overview is presented in Appendix~\ref{sec:appendix:complexities} where the CCs are described separately with examples per identified concern, along with identified mitigation strategies for managing the complexities. Quotes from interviewees of the case organizations are given in Appendix~\ref{sec:appendix:complexities} to provide context for each of the examples.

\begin{table*}[htbp]
% \begin{longtable}{p{0.5cm} p{2cm} p{5.8cm} p{0.5cm} p{1cm}}
\caption{Overview of the contribution complexities, related key concerns, in which case organization they were identified, and examples of similar findings in literature.}
% \label{tbl:contributionComplexities} \\
\label{tbl:contributionComplexities}
\begin{tabular}{p{0.4cm} p{1.7cm} p{6.2cm} p{0.5cm} p{1cm}}
\toprule
\textbf{ID} & \textbf{Contribution complexity} & \textbf{Key concerns} & \textbf{Case \newline Org.} & \textbf{Lit.\newline ref.} \\ \midrule

\multicolumn{5}{@{} l}{\hspace{0.2cm}\textbf{Control-centric Complexities}} \\ \midrule

CC1 &
Impact on value proposition &
$\bullet$ Risk for misalignment between internal and external agendas on OSS with high impact on value proposition. \newline & 1,2 &
~\cite{linaaker2018motivating, linaaker2019community} \\ \midrule

CC2 &
Impact on internal operations &
$\bullet$ Risk for misalignment between internal and external agendas on OSS with high impact on internal operations. & 1,2 &
~\cite{munir2017open, linaaker2019community, linaaker2018motivating} \\ \midrule

\multicolumn{5}{@{} l}{\hspace{0.2cm}\textbf{IPR-centric Complexities}} \\ \midrule

CC3 &
Differentiating functionality &
$\bullet$ Risk of loosing product-based competitive edge. (CaseOrg1-2) \newline
$\bullet$ Risk of loosing process-based competitive edge. (CaseOrg1-2) \newline
$\bullet$ Risk of damaging the business models of existing actors on market. (CaseOrg3) & 1,2,3 &
~\cite{linaaker2018motivating, wnuk2012can,van2009commodification, henkel2006selective,henkel2008champions} \\ \midrule

CC4 &
Commoditi-zation & %a dash hack, should be fixed with latex hyphenation
$\bullet$ Risk for giving away competitive edge or differentiating functionality too early or an alternative solution being accepted before ones' own. (CaseOrg1-2)  \newline
$\bullet$ Risk of damaging the business models of existing actors on market. (CaseOrg3) & 1,2,3 &
~\cite{van2009commodification, bosch2013achieving, linaaker2018motivating} \\ \midrule

CC5 &
Sensitive IPRs  &
$\bullet$ Risk of giving away patented, patentable or in other ways sensitive IPR. & 1,2 &
~\cite{linaaker2018motivating} \\ \midrule

CC6 &
Substitutes &
$\bullet$ Unnecessary cost of contributing if existing alternatives are considered as good or better. & 1,2,3 &
~\cite{linaaker2018motivating} \\ \midrule

CC7 &
License compliance &
$\bullet$ Risk of violating license conditions if software artifact is not contributed. & 1,2,3 &
~\cite{henkel2006selective, linaaker2018motivating} \\\midrule

\multicolumn{5}{@{} l}{\hspace{0.2cm}\textbf{Exposure-centric Complexities}} \\ \midrule

CC8 &
Ethical use &
$\bullet$ Risk of creating negative exposure, hurting brand and public trust. & 3 &
 \\ \midrule

CC9 &
Security threats &
$\bullet$ Risk of exposing security-related vulnerabilities. & 1,2,3 &
~\cite{linaaker2018motivating} \\ \midrule

\multicolumn{5}{@{} l}{\hspace{0.2cm}\textbf{Cost-centric Complexities}} \\ \midrule

CC10 &
Budget and resource constraints &
$\bullet$ Cost for preparing, contributing and maintaining software artifact. & 1,2,3 &
~\cite{kilamo2012proprietary, west2005contrasting, dahlander2008firms} \\ \midrule

CC11 &
Modularity and architecture &
$\bullet$ Technical feasibility to modularize and abstract software artifact. & 1,2,3 &
~\cite{linaaker2019community} \\ \midrule

CC12 &
Code aligment &
$\bullet$ Cost of maintaining internal fork. Misalignment between internal and external development may prevent or complicate future contributions. & 1,2 &
~\cite{munir2017open, linaaker2018motivating} \\ \midrule

\multicolumn{5}{@{} l}{\hspace{0.2cm}\textbf{Community-centric Complexities}} \\ \midrule

CC13 &
External interest &
$\bullet$ Risk of contribution not being accepted or community not being established. & 1,2,3 &
~\cite{linaaker2019community, linaaker2019method} \\ \midrule

CC14 &
Influence in community &
$\bullet$ Low level of influence in the community may prevent or complicate the contribution of a software artifact. \newline
$\bullet$ Target foundation may require a governance model too open which may render in too large scope and loss of control.& 1,2,3 &
~\cite{munir2015open, schaarschmidt2015firms, maenpaa2018organizing, linaaker2019method} \\ \midrule

CC15 &
Community health &
$\bullet$ Not contributing may have a negative impact on the health of the OSS project. & 1,2,3 &
~\cite{linaaker2019community, jansen2012shades, zhou2016inflow} \\  \bottomrule

\end{tabular}
\end{table*}
% \end{longtable}

\subsubsection{Control-centric Complexities}
Control-centric complexities highlight the risk of the development of the software artifact (if released as OSS) heading in another direction than expected by the focal organization and what impact this would have. This is a risk that may be expected as the stakeholder population in a community is heterogeneous with agendas that not always align~\cite{linaaker2019community}. Also if one wants to grow a community, the organization to some extent have to consider the wills and opinions of the community~\cite{laurent2009lessons}. One complexity (CC1) considers the case when the artifact has a close relationship to the organization's value proposition, for example by enabling the organization to sell and deliver its products and services. Another complexity (CC2) considers the case where the artifact may have a more distant relationship to the value proposition but still being of importance in terms of internal operations. Examples highlight OSS projects integrated into the continuous integration tool-chain and used heavily by the product development departments. To address this complexity and the risk of deviating agendas in a community, and organisation can try to build influence in the concerned community through active engagement and contributions.

\subsubsection{IPR-centric Complexities}
Complexities centered around Intellectual Property Rights (IPR) considers what impact it would have on the focal organization's rights if the software artifact is released as OSS. One complexity (CC3) highlights the potential presence of differentiating functionality which could lead to a loss of competitive edge. A recommended practice is to separate between what is commodity and differentiating functionality, for example, through a modular architecture, and contribute underlying parts such as frameworks and libraries. Another complexity (CC4) considers the timing aspect of when the artifact should be released as OSS. It refers to the commoditization cycle where technology moves from an innovative and differentiating state to becoming commodity and general knowledge. One interviewee described it as a \textit{``a trade-off between a competitive edge and burdon''} (I1) implying that by keeping it closed, an organization may guard what it considers differentiating, but as a consequence needs to maintain it by themselves, and at the same time risks that a competing solution gets released and adopted. It is, therefore, important to keep an awareness of potential alternative solutions.

From a public sector perspective, the intention may be the opposite in terms of CC3-4, for example to enable companies to create value based on public assets. This may also help concerned organizations to shift focus from commodity to more value-adding development. However, a risk is that it may damage the business and be viewed as the agency is intending to compete on the market. It may therefore be good to inform concerned organizations in advance and provide them with an opportunity to adapt and enter a dialogue when needed.

Another complexity (CC5) highlights sensitive IPRs in general, which may include patents used in a defensive patent portfolio, as a source of license revenue, or patents belonging to third-party. Hence, a critical review process for the need of the patents and their origin may be motivated in order to weigh these aspects against the potential value that the artifact may produce, if released as OSS.

A frequent scenario reported is that engineers request to release an artifact as OSS when there is already an acceptable alternative available as OSS (CC6). Actively encouraging engineers to research available options is therefore recommended. Education is also seen as a key approach to managing the risk of violating license conditions as these may demand that certain artifacts are contributed or made available as OSS under a certain license (CC7).

\subsubsection{Exposure-centric Complexities}
Exposure-centric complexities highlight ways in how the software artifact may be used that could have a negative exposure of the organization. The artifact may, for example, be used in contexts or for certain purposes that may be considered unethical and not originally intended by the organization (CC8). 

Another risk is that security vulnerabilities are identified before they are fixed which could be used to damage the focal organization and other users of the artifact when available as OSS (CC9). It may therefore be good to pro-actively investigate potential unethical use cases and perform security audits, before releasing any artifact as OSS.

\subsubsection{Cost-centric Complexities}
Constraints in terms of budget and available resources can be an issue as preparing an artifact to be contributed, pushing it through a contribution process, as well as potentially maintaining it once released as OSS has attached costs (CC11). In certain cases the technical feasibility of releasing the artifact may be questioned, as it may be too integrated into an organization's internal software stack, which in turn leads to that the cost may outweigh the potential benefits of releasing it as OSS (CC12). A mitigation can be to, early on, consider the potential of releasing the artifact as OSS and develop the artifact with that in mind, e.g., by using modular architecture and keeping code comments general.

When an artifact relates to an existing OSS project, there is an alternative cost of maintaining the artifact internally and a risk of growing a technical debt by not contributing (CC12). One way to minimize such costs and risks is to keep the internal fork as close as possible to the up-stream OSS project.

\subsubsection{Community-centric Complexities}
One risk is that the external interest is limited for a software artifact intended to be released as OSS (CC13). For existing communities, this would prevent it from being accepted. If creating a new community is considered, but none interested in joining, some of the expected benefits may be missed (e.g., shared maintenance costs). It may therefore be recommended to investigate whether there actually exists external interest, and if the use cases that the software artifact addresses are general enough.

Limited interest from a community may be due to misaligned agendas among its members (CC14). In these cases, it may be important for the focal organization to build up a level of influence through active engagement and contributions, in order to be able to contribute its artifacts.

When an organization is dependent on an OSS project and the community's health may be considered as a risk, it is important to contribute to that community (CC15). Hence, the level of health of the related community should be weighed against other complexities, when deciding if an artifact should be released as OSS.

\section{Discussion}\label{sec:disc}
In this section we discuss the identified contribution considerations, in relation to literature, and with respect to the contexts they were observed.

\subsection{Contribution Objectives}
The expected benefits from sharing a software artifact as OSS is reflected by the identified COs and related key benefits (see Table~\ref{tbl:contributionObjectives}). Objectives can be considered to cover both the idealistic, survival and commercial motivators highlighted by Jansen et al.~\cite{jansen2012shades}. Supportive findings can to a large extent also be found in literature if compared to Section~\ref{sec:rw:rationale}. For example, the idealistically motivated objective CO4 - \textit{Be a good open source citizen}, implying that an organization should respect and understand the needs, governance and culture of the community~\cite{butler2018investigation, nguyen2017coopetition, dahlander2005relationships, dahlander2006man, aagerfalk2008outsourcing}. Further, the commercially motivated cost-saving objectives implied by the extended development workforce (CO12), and related benefits, has also been observed in a number of other studies (e.g.,~\cite{munir2017open, stuermer2009extending, henkel2006selective, lindman2009beyond, olsson2017from, van2009commodification}). Support was also found for the survival, or more strategically, motivated objectives such as CO9 - \textit{Build a software ecosystem}~\cite{west2006challenges, west2008creating}. Some objectives, however, were not reflected among the reviewed literature, e.g., CO5 - \textit{Create price pressure} on third-party vendors and suppliers.

A challenge with the intangible~\cite{sharivar2018business}, or non-monetary benefits~\cite{morgan2014beyond} is that they may be less obvious and harder to measure and track in financial spreadsheets~\cite{morgan2014beyond}, compared to the tangible revenues and monetary benefits. In this study, an attempt is made by presenting key benefits to each objective. Future research could strive to identify and derive suitable and actionable metrics for the different types of benefits, or objectives\footnote{See e.g., https://chaoss.community/ Accessed: \today}. With such metrics in place, contribution requests and strategies can potentially become easier to motivate, execute, follow-up and learn from. Morgan and Finnegan~\cite{morgan2014beyond} highlight how organizations \textit{`` \ldots need to put structures and incentives in place to convert intangible asset capture into something tangible for the [organization]''}.

Objectives further align with the propositions proposed by Munir et al.~\cite{munir2018theory} which states that organizations adopting OSS in their tools and infrastructure setups may benefit through reduced product development costs and time-to-market, as well as increased product and process innovation. Considering Munir et al's~\cite{munir2018theory} classification of why and when organizations adopt and share software as OSS, all three case organizations can be classified as having a proactive approach with the main focus on building symbiotic relationships through their OSS engagements. Similar to Sony Mobile~\cite{munir2018theory}, these organizations can hence be seen as mature players where the use and sharing of OSS are motivated by business rationale.

When comparing the two private case organizations, CaseOrg1 and 2, against the public case organization CaseOrg3, many of the objectives overlapped, although the expected key benefits may differ. For example, regarding CO8, which is focused on creating or replacing an existing industry or community standard, CaseOrg 1 and 2 see value in being able to steer the direction of a community or industry, and potentially disrupting competitors by commoditizing a certain technology~\cite{van2009commodification}. CaseOrg3, on the other hand, views the key benefit as improving competition in the industry or market and potentially forcing the private actors to adapt. Hence, both private and public actors may view the release of OSS and commoditization of the underlying technology as a strategic tool, but in some cases for the opposite reasons.

\subsection{Contribution Complexities}
The contribution complexities presented in Table~\ref{tbl:contributionComplexities} can be viewed from the different perspectives of a contribution strategy. With respect to whether a software artifact or parts of it should be shared as OSS or not, a common concern in literature regards the risk of sharing differentiating or in other ways sensitive IPR ((e.g.,~\cite{wnuk2012can,van2009commodification, henkel2006selective,henkel2008champions,west2006challenges,iivari2008usability}). This risk was also explicitly recognized in the complexities CC3 and CC5. Selective revealing, as labeled by Henkel~\cite{henkel2006selective}, was a recognized practice in both CaseOrg1 and 2, as well as in Sony Mobile~\cite{linaaker2018motivating}. 

In contrast, this was not an issue for CaseOrg3. As a public sector organization, they aim to release what they find most ``differentiating'' in their view, and what can provide the biggest value to their ecosystem, and in the end, the citizens. Their motivation lies in their goals as defined by the government: \textit{``to facilitate and enable matching between job-seekers and employers on the Swedish labor market''}\footnote{https://arbetsformedlingen.se/om-oss Accessed: \today}. However, they do see a balance between raising the bar for private actors in their ecosystem and not hurting their business model.
Although not explicitly found in the coding, this introduces a timing factor, i.e., that the contribution or release of the software artifact is stalled until the private actors have had time to adapt. This addresses the question of when a software artifact should be shared as OSS. 

A related complexity is that of the commoditization process and the concerned software artifact (CC4), which is also brought up in literature~\cite{van2009commodification, wnuk2012can, haruvy2008open, kort2011should, caulkins2013when, linaaker2018motivating}. By being early, an organization can get a first-mover advantage, gain bigger influence, and potentially steer the development, e.g., within a community through a new feature~\cite{linaaker2019community, munir2017open}, or within an industry through a new standard or platform~\cite{west2006challenges} (see e.g.,  contribution objective CO8 and CO9). By being too late, an organization can risk having to adapt to competing solutions or maintain the internal solution~\cite{linaaker2018motivating, ven2008challenges}. Among the case organizations, both CaseOrg1 and 2 experienced this complexity. I1 from CaseOrg1 described it as a \textit{``\ldots trade-off between competitive edge and burdon''}, aligning with other reported observations~\cite{wnuk2012can}.

Lastly, the question of where, i.e., if the software artifact should be contributed to an existing community, or if a new community should be established, touches on several of the complexities, many of which are found across all three case organizations. A common concern among many of the complexities is the need for, or risk of losing, control (e.g., CC1-2, 14). In these cases, if an organization requires a certain level of control on the development direction of the software artifact, the artifact may preferably be released in a community where the organization has a high level of influence on the RE process~\cite{linaaker2019community, schaarschmidt2015firms}, or as a new community~\cite{linaaker2018motivating}. A decisive factor is the external interest for the software artifact (CC13), i.e., whether or not it be accepted within a specific community, or if there is an interest for a new community. In the former case of an existing community, having an existing influence within the community could help to create the interest (CC14)~\cite{munir2017open}. 
The health of the OSS community is another complexity as a community requires active contributions from its members to stay alive. These factors may be considered in a specific \emph{community strategy} as seen from the perspective of the focal firm, which outlines the importance of that community, and what engagement and level of influence that the focal firm wants with respect to the community's requirements engineering process~\cite{linaaker2019community}.

When weighing the options of contributing to an existing community or creating a new one, the earlier may be preferred if possible. As highlighted by CC10, creating a new community may be related to a higher cost, and comes with a number of extra challenges~\cite{kilamo2012proprietary, west2005contrasting}. How this is best done is, however, beyond the scope of this paper, and an interesting topic of further research.

\section{Threats to Validity}
The identified contribution objectives and complexities are the result of a multiple-case study of three case organization over the lapse of two research cycles. To determine the external validity~\cite{runeson2012casestudy} of the objectives and complexities, the characteristics of these organizations need to be considered as they define the problem context~\cite{wieringa2014design} on which the COs and CCs are based.

The three case organizations investigated in this study both have overlapping and distinct characteristics. To some extent, they provide extremes~\cite{flyvbjerg2007five} to each other, while still having resembling characteristics. Considering the way they use and leverage OSS, all three organizations use it for their internal tool and infrastructure setups. In CaseOrg2, the department developing these tools are explicitly studied. CaseOrg1 uses OSS to enable and add value to their hardware devices and as a basis for certain services sold to business-oriented customers. As a public agency, CaseOrg3 differs from CaseOrg1 and CaseOrg2 in that they are not driven by commercial business incentives. Instead, they wish to help private actors focus on more value-adding activities and thereby improve job-matching capabilities for employers and job-seekers. CaseOrg2 however, does not consider themselves as having any product differentiation, compared to CaseOrg1. Following the categorization by Munir et al.~\cite{munir2018theory}, the organizations provide a suitable sample as they all have an awareness for why they share software as OSS and may reflect on the complexities that are present, even in the case of CaseOrg3 who does not have an explicit contribution process in place.

We acknowledge the limitations of case studies and do not claim any statistical generalization~\cite{runeson2012casestudy}. However, we do believe they provide a means to gather deep knowledge of industry practice and rationale in the problem context. By considering the case organizations' characteristics, the reader can put the presented contribution objectives and complexities as well as the organizations' rationale and concerns for sharing software as OSS into context. Through analytical generalization (cf. analogical inference~\cite{wieringa2014design}), results from this study can then be extended to cases with similar characteristics within a similar context~\cite{runeson2012casestudy}. Both similarities and dissimilarities between the source and target cases should be thoroughly analyzed~\cite{ghaisas2013generalizing}. In Section~\ref{sec:caseOrgs}, we provide a general description of each of the case organizations, as well as of their specific contribution processes and examples of OSS projects that they are contributing to or have released as OSS. Quotes from interviewees (see examples in Appendices \ref{sec:appendix:objectives} and \ref{sec:appendix:complexities}) are used to describe the contribution objectives and complexities, and to provide further contextual factors that may otherwise risk being lost in the reporting of the research if abstracted by the researcher.

A threat in regards to reliability relates to that only the first author was involved in the data gathering and analysis process of the study. To minimize the risk of misinterpretations, member-checking~\cite{easterbrook2008selecting} was performed by presenting interview summaries to key interviewees at the case organizations. During the coding process, the inductive approach infused a constant comparison between new data and existing codes, enabled by audit-trails being kept~\cite{runeson2012casestudy}. Cross-case analysis~\cite{seaman1999qualitative} was also performed as codes identified in one case were reiterated in the transcripts from the other cases, after which a final coding scheme assembled. As can be noticed in table~\ref{tbl:coding}, there was saturation in the number of emerging contribution objectives and complexities. Triangulation~\cite{runeson2012casestudy} with archival data was also performed by cross-checking coding schema and basing interview questionnaires on contribution request forms from seven organizations.

After the final coding was performed in the second research cycle, the contribution objectives and complexities were presented and discussed with I1 from CaseOrg1, I7 and I8 from CaseOrg2 and I15 from CaseOrg3. This kind of static~\cite{gorschek2006model} or descriptive validation~\cite{hevner2004design} is a useful approach in the artifact validation phase to gain feedback and refine a research artifact before it is transferred or implemented in a real-world problem context~\cite{wieringa2014design, gorschek2006model}.

I1 from CaseOrg1 confirmed identified contribution objectives and complexities. I1 added further facets to e.g. CC6, that patents may also provide a source of license revenues in other cases which should also be considered. I1 found the objectives and complexities useful and that \textit{``[they] really fit into [CaseOrg1's] strategy which is, we are trying to streamline the entire [contribution] process so that we are asking the right questions and we want to get to a point where we are able to approve everything online and don't need to have a meeting. Because if these questions are answered correctly then we should be able to even have an AI/ML-based algorithm which says, the risk is 10\% so it is OK to approve it''}. I1 also adds that a contribution may be \textit{``more things than just code''}, exemplifying evangelizing, technical writing, writing bug reports, sharing of knowledge and experience, as well as test cases and design documentation. We agree with I1 that a contribution may be many things, but choose to limit the scope to the sharing of software artifacts as OSS. We view these artifacts as internally developed software functionality of different size and complexity, e.g., bug-fixes and features to frameworks, projects, and products, including e.g., related test cases and documentation. Other activities we believe should be covered by a community strategy that specifies how an organization should engage with a specific OSS community~\cite{linaaker2019community}.

I7 and I8 from CaseOrg2 found the contribution objectives and complexities valuable and saw value in comparing with other organizations. They believed that the objectives and complexities can be used as an eye-opener to those within the organization that are skeptic to OSS. Both interviewees agreed with and verified the relevance of all of the identified objectives and complexities. They also added support for CC5 and CC15 and refined the description of CO7.

I15 from CaseOrg3 confirmed the identified objectives and complexities, while also adding support for CC4 and CC5. In regards to the latter complexity, I15 highlights how it also connects to CC4 and how they strive to share software artifacts that are differentiating in order to help businesses focus on more value-adding activities.

In conclusion, we believe that the identified contribution considerations are relevant to the studied case organizations, but we also believe that the findings have relevance beyond these organisation, while that of course depends on problem context to which these findings are transferred. The validation of the completeness of the set of contribution considerations, is a matter for continued research, but our investigation of existing literature has not indicated any significant omissions.

\section{Conclusion}\label{sec:conclusions}
In this study, we aim to help organizations in deciding if a software artifact (e.g., feature, framework or project) should be released as OSS, and, if so, when and where. For a specific artifact, the ''\textit{what}''-question regards if the artifact should be contributed in full or kept closed, or if certain parts can be contributed under certain conditions. The ''\textit{when}''-question regards the timing of the release. Finally, the ''\textit{where}''-question regards whether the artifact should be contributed to one of many existing OSS communities or if a new community should be established. Answers to these questions may be valuable input to the development of a \textit{contribution strategy} for the concerned software artifact.

To support organizations in creating effective contribution strategies, we conducted a multiple-case study at three software-intensive organizations using an iterative approach spanning over two research cycles. A set of 27 contribution considerations, divided into 12 objectives and 15 complexities, were identified based on an inductive and iterative coding of 20 interviews across the three case organizations. %By identifying relevant objectives and complexities, an organization can create contribution strategies based on its specific context and business goals.

Contribution objectives highlight opportunities for 1) improving reputation towards community, customers, partners, and current and future employees, 2) managing suppliers through price pressure and outsourcing, 3) managing partners and competitors through standardization efforts and ecosystems, and 4) exploiting externally available knowledge and resources through open innovation and extended development resources. 

Contribution complexities focus on 1) risk of losing control of the development of business-critical software in concerned communities, 2) risk of giving away differentiating IPR that provides a competitive advantage, 3) risk of unintentionally exposing unethical use-cases and security vulnerabilities, 4) costs of abstracting and generalizing software artifact, pushing through contribution process and potentially maintaining once contributed, and 5) difficulties in deciding where the artifact should be contributed based on external interest, potential need for influence, and community health.

Future work should look to generalize these identified contribution considerations beyond the problem context of the three investigated case organizations. In order to arrive at a framework that can be used by software-intensive organizations when setting up their contribution strategy decision process, further validation and generalization is need. Specific research focus is also needed to investigate possible relationships between contribution objectives and contribution complexities. In addition, research attention could be given to identifying and formalizing metrics that may be used for quantifying the potential benefits proposed in the objectives, as well as the risks and costs implied by the complexities. Such metrics could help organizations arrive at key performance indicators, such as a Return-on-Contribution, which may help to make decisions comparable, measurable, and easier to motivate. This could further help organizations in automating their contribution processes and thereby potentially lowering the threshold for developers to contribute, and also, hopefully, help the organization to more effectively reach their contribution objectives.

\begin{acknowledgements}
We would like to thank the anonymous reviewers for their constructive feedback from which the paper has benefited greatly. Also, we would like to thank the interviewees and case organizations for participating in this study. Without their participation the study would not have been made possible.
\end{acknowledgements}

\begin{appendices}
% \titleformat{\section}{\bfseries}{Appendix \thesection.}{0.5em}{}
%\titleformat{\subsection}{\normalfont\itshape}{\thesubsection.}{0.5em}{}
%\titleformat{\subsubsection}{\normalfont\itshape}{\thesubsubsection.}{0.6em}{}

\section{Questionnaire - First Research Cycle}\label{sec:appendix:secondCycle}

\noindent\textbf{Demographics:}
\begin{itemize}
\item What is your job title?
\item How many years of experience do you have in your current and similar roles?
\item Could you, in short, describe your daily work and responsibilities?
\item Could you, in short, describe your experience in working with OSS communities?
\end{itemize}

\noindent\textbf{Contribution strategy:}
\begin{itemize}
\item Do you, in any way, consider what you contribute and reveal as open source? If yes, how? Is it formalized in any way? How could this be improved/otherwise done? What connections do you see to a company's ROI and competitive edge?
\item How would commoditization affect what you contribute and not? How would it affect the timing of when you contribute?
\item How would a feature's profit for the company affect what is contributed and not? How would it affect the timing of when you contribute? How could this aspect be judged or quantified?
\item How would it affect if the feature or the knowledge and technology behind it is hard to acquire or control? How could this aspect be judged or quantified?
\item How would you reason in regard to if and when to contribute, given that a feature results in a:
    \begin{itemize}
    \item high profit for your company, and is critical to maintain control over?
    \item low profit for your company, and is critical to maintain control over?
    \item high profit for your company, and not critical to maintain control over?
    \item low profit for your company, and is not critical to maintain control over?
    \end{itemize}
\item What other aspects would affect your decisions?
\item How would your decisions affect how you engage and invest in the community where the feature is to be contributed to?
\item How are policies or decisions regarding what to contribute communicated today? How could this be improved/otherwise done?
\item Is there anything I have missed that you would like to pick up on? Or anything else that you would like to talk about?
\end{itemize}

\section{Questionnaire - Second Research Cycle}\label{sec:appendix:thirdCycle}

\noindent\textbf{Demographics:}
\begin{itemize}
\item What is your job title?
\item How many years of experience do you have in your current and similar roles?
\item Could you, in short, describe your daily work and responsibilities?
\item Could you, in short, describe your experience in working with OSS communities?
\end{itemize}

\noindent\textbf{Contribution Objectives:}
\begin{itemize}
\item What reasons do you see to share your internally developed software as OSS? What can your organization gain out of this? How can you measure it?
\item What reasons do you see in relation to competition and collaboration with external parties? Who would it benefit or hurt?
\item What effect can you have on a third party by releasing software as OSS? For example, competitors or suppliers of similar proprietary products?
\item Do you consider if there is a potential to create a new standard? How can such potential be measured?
\item Are there any type of software that should be shared for different reasons? How would these types be characterized?
\end{itemize}

\noindent\textbf{Contribution Complexities:}
\begin{itemize}
\item What aspects should you take into consideration in the decision of sharing software as open source or not?
\item What consideration should be taken to how software relates to your value proposition and how it affects your revenues? For example, if the software is shipped with your product or if it is used to build or deliver your product/service?
\item How is a decision affected if the software contains differentiating parts?
\item How is a decision affected if the software contains patents or parts that could be patented?
\item Is there any type of sensitive information within the software that could complicate a contribution?
\item How should any competitive advantages be taken into consideration?
\item How do timing and commoditization affect such aspects and any future decisions? How do you consider the risk of competing solutions being released before yours?
\item How do you consider internal restrictions in terms of budget and resources? What costs do you see related to releasing software as OSS? How do these costs affect the decision?
\item Do you see any other business related aspects, impediments or risks that should be taken into consideration?
\item What consideration should be taken in regards to how the software is used and integrated into your organization and your product/services?
\item What consideration should be taken to security-related aspects?
\item How do you consider the software's generalizability and potential to extend?
\item Do you see any other technical aspects, impediments or risks that should be taken into consideration?
\item How is a decision affected in terms of the need to control the development of the software?
\item How do you consider existing alternatives (OSS and proprietary)?
\item What alternatives to you see if there is a shallow external interest from a community?
\item What factors affect your decision when choosing to contribute to an existing community or creating a new community?
\item How do you consider the health and sustainability of a community?
\end{itemize}

\section{Findings on Contribution Objectives} \label{sec:appendix:objectives}
%\begin{small}

\subsection{Reputation-centric Objectives}

\subsubsection{CO1 - Prove skill and influence:}

\begin{itemize}
    \item[\textbf{Description:}] Being an active contributor in a community can help to build a reputation of an organization as technically skilled and influential in the community~\cite{munir2015open, ven2008challenges}.
    \item[\textbf{Benefit:}] Improved trust towards customers.
    \item[\textbf{Example (CaseOrg1):}] \textit{``We basically realized that we were so dependent on [OSS project] and that it was the selling point of our product, so we needed to demonstrate for customers that we were one of the core contributors. It was a selling point in the marketing brochure''} (I1).
    \item[\textbf{Benefit:}] Improved trust towards partners.
    \item[\textbf{Example (CaseOrg2):}]  I7 reports how being an active contributor in a community can help to promote CaseOrg2 in new channels and build up new partnerships.
\end{itemize}

\subsubsection{CO2 - Increase transparency:}

\begin{itemize}
    \item[\textbf{Description:}] By being transparent in how an organization performs certain actions, they can build trust among customers and potentially avoid allegations of wrongful doing.
    \item[\textbf{Benefit:}] Improved trust among customers.
    \item[\textbf{Example (CaseOrg1):}] A tool was developed and open sourced to measure the speed of their services. By keeping the project open, the organization could be transparent in how they performed the measurement.
    \item[\textbf{Benefit:}] Improved trust among the public.
    \item[\textbf{Example (CaseOrg3):}] From a public sector perspective, CaseOrg3's aim is in a similar manner to create transparency towards the public and thereby earn their trust, \textit{``this is what you get for your tax money''} as put by I13.
\end{itemize}

\subsubsection{CO3 - Improve employer branding:}

\begin{itemize}
    \item[\textbf{Description:}] Working with OSS projects can help an organization to both retain existing and attract new engineers~\cite{munir2015open}.
    \item[\textbf{Benefit:}] Attraction of talented employees.
    \item[\textbf{Example (CaseOrg1):}] \textit{``It is important to people and I think a positive employment characterization that you get to engage with open source communities and that the company does release something as open source projects. I think that is a big selling point that people are looking for in whom they want to work for as an engineer''}.
    \item[\textbf{Benefit:}] Retention of existing employees.
    \item[\textbf{Example (CaseOrg1\&3):}] Besides being allowed to engage with an external community with the intrinsic rewards that follow, highlighted by I19 as \textit{``the fun parts''}, working with OSS offers engineers the opportunity build their own transparent portfolios. As stated by I6, \textit{``[the engineers] view it as a core part of their career development and in a way, they can kind of be recognized for that in a very different way than they could if they were simply working on internal or commercial software''}.
\end{itemize}

\subsubsection{CO4 - Be a good open source citizen:}

\begin{itemize}
    \item[\textbf{Description:}] Contributing to an OSS project may for some be ideologically motivated~\cite{jansen2012shades}.
    \item[\textbf{Benefit:}] Idealistic satisfaction among employees.
    \item[\textbf{Example (CaseOrg1):}] \textit{``I think it is from one point of view the right thing to do, because you are taking advantage of this set of code, and so you should, while you are taking advantage of it, you should also be contributing back [from] a good citizen point of view''}. I5 further adds that a reason \textit{``may be that the developers are pro-OSS and so they are just like, this isn't secret sauce, so just sort philosophically we want to contribute this to the commons because maybe someone will take advantage of it. There's a good citizen aspect to it''}.
\end{itemize}

\subsection{Supplier-centric Objectives}

\subsubsection{CO5 - Create price pressure:}

\begin{itemize}
    \item[\textbf{Description:}] Releasing a project as OSS can be a way to put price pressure on third-party vendors and suppliers of proprietary solutions.
    \item[\textbf{Benefit:}] Lower subscription costs of produced products and services.
    \item[\textbf{Example (CaseOrg1):}] As a reaction to an expensive proprietary infrastructure solution, CaseOrg1 developed an internal replacement and open sourced it as \textit{``a competitive advantage in the negotiations against [the supplier of the proprietary solution]''} (I4) in order to \textit{``drive the cost of doing business down''} (I5).
    \item[\textbf{Benefit:}] Lower prices on tenders.
    \item[\textbf{Example (CaseOrg3):}] From a public sector perspective, by releasing software as OSS, and requiring tenders to build on it, competition in the tender process is improved as smaller actors enabled to participate in a process \textit{``otherwise set-up to benefit large firms''} (I12).
\end{itemize}

\subsubsection{CO6 - Outsource infrastructure operation:}

\begin{itemize}
    \item[\textbf{Description:}] Sharing software as OSS can be seen as a way to not just outsource the development of a software project~\cite{aagerfalk2008outsourcing}, but also the operation of it.
    \item[\textbf{Benefit:}] Internal focus on activities related to core business.
    \item[\textbf{Example (CaseOrg1):}] As explained by I5, \textit{``We can get to a point where we can actually outsource the operations of this [software] ... via an open source path, which then frees up this team to do other stuff''}.
\end{itemize}

\subsection{Strategy-centric Objectives}

\subsubsection{CO7 - Collect data:}

\begin{itemize}
    \item[\textbf{Description:}] By sharing e.g., algorithms as OSS, others can generate data which can then be used to improve artificial intelligence and machine learning initiatives.
    \item[\textbf{Benefit:}] Improved machine learning and artificial intelligence algorithms.
    \item[\textbf{Example (CaseOrg1):}] I1 explains \textit{``The Machine learning team has often come to us saying, I want to open source this algorithm because I wanna learn how it grows, and we alone don't have enough data to feed and we need the world to feed it data''}.
    \item[\textbf{Benefit:}] Creation of solutions based on Open Data sources.
    \item[\textbf{Example (CaseOrg3):}] I15 exemplifies how the gathering of job-ads can help create an automated review-function which could classify if an add is discriminatory or not.
\end{itemize}

\subsubsection{CO8 - Standardize a solution:}

\begin{itemize}
    \item[\textbf{Description:}] By having an OSS project or contribution adopted and accepted as a standard solution, an organization can potentially replace existing proprietary and community solutions~\cite{lindman2009beyond, jansen2012shades, henkel2006selective, west2003open}.
\item[\textbf{Benefit:}] Force competitors to adapt and steer market or community according to internal agenda.
    \item[\textbf{Example (CaseOrg1):}] A project was developed and open sourced in order to replace proprietary solution as \textit{``a way to get away from a commercial vendor and open up the access to the data collected by the vendor, but also to be able to influence and control [the project's] direction''} (I6). As further explained by I1, \textit{``by putting it out there you can influence it and thereby where the market is heading''}, potentially steering the competition in the direction most beneficial to the firm.
    % \item[\textbf{Example (CaseOrg2):}] In another example, CaseOrg2 wanted to replace an existing plugin related to an OSS project they are dependent on. If successful, it would \textit{``mean less of a burden when upgrading to new releases''} and \textit{``not having to maintain an internal fork''} (I8).
    \item[\textbf{Benefit:}] Improve competition and enable more value-adding development on market.
    \item[\textbf{Example (CaseOrg3):}] From a public sector perspective, CaseOrg3 views the benefit as \textit{``improving competition''} (I13) and \textit{``allowing more to join the market''} (I15). I15 continues, \textit{``I think we are raising the bar for the entire market, allowing everyone to level up and stop focus on base infrastructure''}.
\end{itemize}

\subsubsection{CO9 - Build a software ecosystem:}
\begin{itemize}
    \item[\textbf{Description:}] Software released as OSS can serve as a technological platform around which a software ecosystem may form with firms who can start to collaborate and interact through a shared market of software and services.
    \item[\textbf{Benefit:}] Enable and stimulate creation third party applications and services.
    \item[\textbf{Example (CaseOrg3):}] From a public sector perspective, CaseOrg3 is aiming to create a software ecosystem by contributing internal software artifacts as new OSS projects along with Open Data sources. Based on the platform, constituted by the OSS projects and Open Data sources, private organizations and firms on the labor market can create new and \textit{``hopefully better market-adapted solutions''} (I13). I17 adds \textit{``I think there is an interest and a need for everybody within the sector to collaborate and to help each other and to make sure that the right individuals land the right opportunities. So what we are trying to gain is literally a collaboration [...] with others to provide better services to society''}.
\end{itemize}

\subsubsection{CO10 - Improve partner collaboration:}

\begin{itemize}
    \item[\textbf{Description:}] Sharing e.g., tools and infrastructure projects as OSS could potentially act as a complement and improve collaboration between CaseOrg2 and other actors within CaseOrg2's software ecosystem that is connected to its core-business of embedded devices.
    \item[\textbf{Benefit:}] Increased value of software ecosystem.
    \item[\textbf{Example (CaseOrg2):}] As put by I7, \textit{``It would be easier to collaborate around the same stack... Also, by offering boilerplates to our partners we help them to do the right thing from start''}. I8 adds, \textit{``I think, would [CaseOrg2] be more open with its tools and libraries, this would be appreciated by, and a way to come closer to our third-party developers''}.
\end{itemize}

\subsection{Engineering-centric Objectives}

\subsubsection{CO11 - Open up innovation process:}

\begin{itemize}
    \item[\textbf{Description:}] Collaboration with OSS communities can be seen as a case of Open Innovation~\cite{munir2015open}, allowing an organization to extend its R\&D and innovation capabilities and initiate new collaborations in an open and transparent way.
    \item[\textbf{Benefit:}] Accelerated innovation process.
    \item[\textbf{Example (CaseOrg1):}] I6 describes it as \textit{``Open source is in many cases about doing external R\&D in a way. You are leveraging external groups to do things, and it is often the case that many companies don't really fund medium to longer term R\&D in-house anymore... You get the network effect of so many more participants''}.
    \item[\textbf{Benefit:}] Creation of more and better market-oriented solutions
    \item[\textbf{Example (CaseOrg3):}] Open collaboration may also be seen as an opportunity to \textit{``open up the requirement engineering process''} (I15) and create \textit{``faster feedback-loops, and to become more reactive and compatible to the needs of the market''} (I12).
\end{itemize}

\subsubsection{CO12 - Extend development resources:}

\begin{itemize}
    \item[\textbf{Description:}] Using and combining the development resources of an OSS community with an organization's internal, software development may potentially be accelerated and extended as e.g., features are implemented and bugs corrected~\cite{munir2017open}.
    \item[\textbf{Benefit:}] Focus on more value-adding development.
    \item[\textbf{Example (CaseOrg1):}] I6 describes it as \textit{``the more that you can push things down to commodity, the easier it becomes, and cheaper, and then you can keep focusing more up the stack. Keep focusing on the next new feature''}.
    \item[\textbf{Benefit:}] Accelerated development.
    \item[\textbf{Example (CaseOrg1):}] \textit{``[Y]ou almost get this acceleration of innovation on that platform. It supercharges [the platform] in a way and you are getting, maybe it is us and five other actors, maybe even competitors, and we are all contributing back to things that we are finding. It is a better product for everybody. So I think that accelerates the development process''} (I6).
    \item[\textbf{Benefit:}] Lower maintenance cost.
    \item[\textbf{Example (CaseOrg3):}] \textit{``The code is actually a cost and when you realize that, you will see that it is better to share that cost than keeping it to yourself''} (I12).
\end{itemize}

% In the case of CaseOrg2, the external workforce provides an important leverage as the organization generally have \textit{``limitations in terms of time and resources''} (I7). From a public sector perspective, I13 views OSS as a way to generate a \textit{``higher value in return for the tax money invested''} compared to the closed option.

%\end{small}

\section{Findings on Contribution Complexities} \label{sec:appendix:complexities}

%\begin{small}

\subsection{Control-centric Complexities}

\subsubsection{CC1 - Impact on the value proposition:}

\begin{itemize}
    \item[\textbf{Description:}] Software artifacts that have a close connection to an organization's value proposition and its revenue stream may warrant special attention in order to determine any potential costs or negative risks that may be implied by contributing the artifact.
    \item[\textbf{Concern [CaseOrg1\&2]:}] Risk for misalignment between internal and external agendas on OSS with high impact on value proposition.
    \item[\textbf{Example [CaseOrg2]:}] In the case of CaseOrg2, the department studied in this paper focuses on infrastructure and tools-projects which is used by CaseOrgs2's product development teams. The software developed and maintained by the department can hence be considered to have an indirect connection to CaseOrg2's value proposition and revenue streams, or as put by I7, \textit{``not even close to core business''}. This provides the department with a \textit{``much less restrictive process than what applies for core business''} (I7).
    \item[\textbf{Example [CaseOrg1]:}] CaseOrg1 develops and maintains similar types of projects, but also those that have a stronger connection to the organization's value proposition and revenue streams. For example, one project that they have released is a Linux-based operating system embedded in hardware devices which enables the organization the deliver its service offerings to its customers. Another project released by the organization makes up a pivotal part of their infrastructure, also enabling the delivery of its service offerings, but also the possibility to offer the infrastructure as a service to business customers, including competitors. In both examples, the projects are considered as commodity but of strategic importance why the organization needs to be \textit{``actively involved and be able to steer the direction and make sure that as that project evolves that it continues to meet our needs and to evolve in a certain way''} (I5).
    \item[\textbf{Mitigation strategy:}] Build influence on OSS community's RE process to enforce internal agenda.
\end{itemize}

\noindent\textbf{CC2 - Impact on internal operations:}

\begin{itemize}
    \item[\textbf{Description:}] Even though a software artifact may have a loose connection to a organization's value proposition and revenue stream, it may still be of strategic importance internally. By releasing the artifact as OSS there is a risk of control being too diluted due to other stakeholders' interests.
    \item[\textbf{Concern [CaseOrg1\&2]:}] Risk for misalignment between internal and external agendas on OSS with high impact on internal operations.
    \item[\textbf{Example [CaseOrg2]:}] \textit{``We are extremely dependent on [OSS project] as we have built our whole continuous integration tool-chain on it, same goes for [OSS project]. Hence, we need to be active so that we can affect the direction on the projects, otherwise, it could become extremely costly for us. Better tools enable us to make better products''} (I8).
    \item[\textbf{Example [CaseOrg1]:}] I1 phrases it as, \textit{``Which projects are we actively using as a company that we are so dependent upon for our success that we need to be at the table? You can do a survey on the company and ask, what open source infrastructure are you using, and how critical is it to your success? And how are you engaging today? And what is missing in your engagement?''}. I4 adds, \textit{``We do need to influence [the OSS project's] roadmap because as the project continues to evolve, and we are such a huge user of it, we still need to drive its roadmap quite a bit to be able to get maximum value from it''}.
    \item[\textbf{Mitigation strategy:}] Build influence on OSS community's RE process to enforce internal agenda.
\end{itemize}

\subsection{IPR-centric Complexities}

\subsubsection{CC3 - Differentiating functionality:}

\begin{itemize}
    \item[\textbf{Description:}] Giving away software artifacts that provide product differentiation or other types of competitive edge is a common fear among firms, but also an opportunity and a balancing act for public organizations.
    \item[\textbf{Concern [CaseOrg1\&2]:}] Risk of loosing product-based competitive edge.
    \item[\textbf{Example [CaseOrg1]:}] I3 explains, \textit{``when I look at what we've approved so far, practically 93 percent of what's coming in front of the committee we approve. The 7 percent has been the things that are core to our business, or something that is a competitive advantage''}. This may be a consequence of CaseOrg1's work to actively encourage its developers to separate between differentiating and commodity functionality as explained by I6, \textit{``These things are really basic functionality, that is ok, we are going to share that, that is part of the core platform. But there are maybe some parts where you can on a modular basis extend it or do integrations with internal systems in order to not give away the secret sauce, the differentiation''}.
    \item[\textbf{Mitigation strategy:}] Identify and separate between differentiating and commodity functionality when possible.
    \item[\textbf{Concern [CaseOrg1\&2]:}] Risk of loosing process-based competitive edge.
    \item[\textbf{Example [CaseOrg2]:}] For the department studied in CaseOrg2, a bigger concern relates to the risk of giving away process-based competitive edge, as explained by I7 \textit{``Of course we have tools that can provide us with a competitive edge''}. Faster development pace and better quality assurance are two areas highlighted by I9.
    \item[\textbf{Mitigation strategy:}] Identify and separate between differentiating and commodity functionality when possible.
    \item[\textbf{Concern [CaseOrg3]:}] Risk of damaging the business models of existing actors on market.
    \item[\textbf{Example [CaseOrg3]:}] Concerning CaseOrg3, they prioritize releasing software artifacts that have the potentially highest differentiating value for actors on the market. However, as expressed by I15, \textit{``It is a balancing act. There will be cases where we will open up stuff that some make a living on. Actors will have to innovate their business models and adapt. Our aim is to improve competition, not hurt it''}.
    \item[\textbf{Mitigation strategy:}] Inform actors and provide possibility to adapt in advance.
\end{itemize}

\subsubsection{CC4 - Commoditization:}

\begin{itemize}
    \item[\textbf{Description:}] The timing of when to release something as OSS can be a complex issue. As I1 highlights, \textit{``There is a trade-off between a competitive edge and burdon''}. I.e., while keeping the software closed may provide a competitive advantage, it may cost in terms of maintenance and support.
    \item[\textbf{Concern [CaseOrg1\&2]:}] Risk for giving away competitive edge or differentiating functionality too early or an alternative solution being accepted before ones' own.
    \item[\textbf{Example [CaseOrg1]:}] I1 describes as a \textit{``waiting game... [Some] may want to say - I want to hold on to this feature longer - because they don't want to make it a level playing field for everybody, or they may take certain features and say, - I think we need to get it out there as fast as possible, because we heard that competitor X is working on putting this into the pipeline... So I think it could be both ways, there could be some features where we say, No, let us hold on to this until we get a first-mover advantage on the market using it, then once we have a significant advantage, then it is ok to commoditize it and put it out there''}.
    \item[\textbf{Mitigation strategy:}] Maintain awareness in community and industry about potential alternative solutions.
\item[\textbf{Concern [CaseOrg3]:}] Risk of damaging the business models of existing actors on market.
    \item[\textbf{Example [CaseOrg3]:}] CaseOrg3 also considers the commoditization process as an important aspect, but as highlighted in CC4 they strive to share software artifacts that are differentiating in order to help businesses focus on more value-adding activities.
    \item[\textbf{Mitigation strategy:}] Inform actors and provide possibility to adapt in advance.
\end{itemize}

\subsubsection{CC5 - Sensitive IPRs:}

\begin{itemize}
    \item[\textbf{Description:}] Sensitive IPs or patents is not limited to functionality that provides a competitive edge in terms of a product or process (see CO4). They can also constitute pieces in a defensive patent portfolio, serve as a revenue source of license subscriptions, or belong to a third-party.
    \item[\textbf{Concern [CaseOrg1\&2]:}] Risk of giving away patented, patentable or in other ways sensitive IPR.
    \item[\textbf{Example [CaseOrg1]:}] For CaseOrg1, patents can be viewed as a competitive edge even in cases where they do not cover functionality that is actively used. As I1 explains, \textit{``We live in a very litigious environment, which means, as a company we are in an industry where there are lots of lawsuits against each other, so we use our patent portfolio in a defensive way. So there is a drive to build a patent portfolio which is why the patent office encourages people to seek patents, because the bigger and better the patent portfolio, the more we can defend ourselves''}. Hence, a common question that is asked in CaseOrg1 is \textit{``can and should we patent this?''} (I1). I1 also adds that in other organizations certain patents may provide license revenue, which should also be weighed against the potential value of sharing the patent as OSS.
    \item[\textbf{Example [CaseOrg1]:}] In the case of CaseOrg2, as they are orchestrating a larger software ecosystem around its products, a question they ask is if there are any patents or IPRs that belong to one of their partners.
    \item[\textbf{Mitigation strategy:}] Critically review need and origin of patents.
\end{itemize}

\subsubsection{CC6 - Substitutes:}

\begin{itemize}
    \item[\textbf{Description:}] If there are existing alternatives to the software artifact available, it may be questioned why the artifact should even be considered.
    \item[\textbf{Concern [CaseOrg1,2\&3]:}] Unnecessary cost of contributing if existing alternatives are considered as good or better.
    \item[\textbf{Example [CaseOrg1]:}] according to I3, \textit{``One of the questions we ask in the open source contribution form is, is there an existing project that does the same thing or is this one something unique and new? We've had people that, say, I wrote an HTTP client, and I want to open source it, that question goes to that, hey, there are plenty HTTP clients, why are you writing another one?''}. However, sometimes it may be motivated if there is a strategic intent to replace an existing solution in a community or an industry standard.
    \item[\textbf{Mitigation strategy:}] Educate developers to research substitutes and motivate why artifact still should be released as OSS.
\end{itemize}

\subsubsection{CC7 - License compliance:}

\begin{itemize}
    \item[\textbf{Description:}] In cases where the software extends or integrates with an OSS project, the OSS license of that project needs to be reviewed and respected. Copyleft and restrictive licenses may require that the artifact is contributed.
    \item[\textbf{Concern [CaseOrg1,2\&3]:}] Risk of violating license conditions if software artifact is not contributed.
    \item[\textbf{Example [CaseOrg2]:}] I7 highlights, \textit{``we must be compliant with the licenses we have in our source code''}.
    \item[\textbf{Mitigation strategy:}] Implement compliance programs, educate engineers and automated license-checking.
\end{itemize}

\subsection{Exposure-centric Complexities}

\subsubsection{CC8 - Ethical use:}
\begin{itemize}
    \item[\textbf{Description:}] By releasing a software artifact as OSS, anyone is allowed to use it under the same conditions provided by the OSS license. Hence, a risk is that the software may be used for purposes not originally intended.
    \item[\textbf{Concern [CaseOrg3]:}] Risk of creating negative exposure, hurting brand and public trust.
    \item[\textbf{Example [CaseOrg3]:}] I18 sees a a risk of \textit{``cases where we would not be very proud of as a public agency''} and asks \textit{``how far does the responsibility of CaseOrg3 as a public agency stretch?''}
    \item[\textbf{Mitigation strategy:}] Investigate potential alternative use cases of software artifact.
\end{itemize}

\subsubsection{CC9 - Security threats:}

\begin{itemize}
    \item[\textbf{Description:}] Releasing a software artifact as OSS may pose a security threat by exposing unknown vulnerabilities present in its source code.
    \item[\textbf{Concern [CaseOrg1,2\&3]:}] Risk of exposing security-related vulnerabilities.
    \item[\textbf{Example [CaseOrg2]:}] At CaseOrg2, \textit{``the security department has a positive view on open source as you get more eyes on the code''} (I11). However, they still have a careful review process in place as they do not wish to expose any potential back-doors that could lead to the organization's hardware-based products.
    \item[\textbf{Example [CaseOrg3]:}] At CaseOrg3, I18 highlights that \textit{``it is the data that is considered valuable and needs protection''} which is rather done by \textit{``proper key management''} than keeping any related software closed.
    \item[\textbf{Mitigation strategy:}] Include security audits in contribution process.
\end{itemize}

\subsection{Cost-centric Complexities}

\subsubsection{CC10 - Budget and resource constraints:}

\begin{itemize}
    \item[\textbf{Description:}] Contributing a software artifact always comes with a cost, both in terms of preparing, contributing and potentially maintaining the software artifact as OSS.
    \item[\textbf{Concern [CaseOrg1,2\&3]:}] Cost for preparing, contributing and maintaining software artifact.
    \item[\textbf{Example [CaseOrg2]:}] Abstracting, modularizing and generalizing an artifact may prove an expensive effort compared to projects that are developed with the intention from the start. As put by I8, \textit{``if we build something that is tailored to and dependent on internal infrastructure, it is most often no idea to contribute it. In most cases, we can modularize and generalize it, but the cost can get really high, so it is a matter of if it is worth it or not''}. I9 further mentions that there \textit{``is a lot of hidden costs''} that are connected to the contribution process, such as \textit{``scrubbing''} or cleaning the source code and its version history of sensitive or unnecessary comments and information. Other costs include going through the contribution process of the related community, or creating a community if the software artifact is released independently of any existing community. In the latter case, much more resources are required long-term both in terms of managing the community, but also maintaining and leading the software development within the community. As highlighted by I10, \textit{``we prefer contributing to existing communities because it is expensive taking on the role of a maintainer in a larger project''}.
    \item[\textbf{Example [CaseOrg3]:}] CaseOrg3 recognizes the costs implied by sharing a software artifact as OSS as a concern, but sees it from a long-term perspective. I15 asks, \textit{``What is the need of the labor market? If there is a potentially positive outcome, then I think we are prepared to spend the money needed''}.
    \item[\textbf{Mitigation strategy:}] Develop software artifacts as if they were intended to be released as OSS from start, separating commodity and differentiating functionality. Also, ask \textit{``who is going to be the owner of this repo, and have you allocated the time to maintain it, and has your boss approved your time budget?''} (I1).
\end{itemize}

\subsubsection{CC11 - Modularity and architecture:}

\begin{itemize}
    \item[\textbf{Description:}] In certain cases, it may be that the software artifact is too embedded in internal infrastructure, which makes it infeasible to modularize the artifact.
    \item[\textbf{Concern [CaseOrg1,2\&3]:}] Technical feasibility to modularize and abstract software artifact.
    \item[\textbf{Example [CaseOrg]:}] I1 stresses that \textit{``companies should start projects with the goal that it will be open one day, then they could architect it right from the start with generality and modules that can be contributed''}.
    \item[\textbf{Mitigation strategy:}] Develop software artifacts as if they were intended to be released as OSS from start, separating commodity and differentiating functionality. An option may be to contribute the \textit{``design document or the blueprint of the project itself so someone else can create it''}, i.e., documentation that in some way captures the knowledge from the underlying software artifact.
\end{itemize}

\subsubsection{CC12 - Code alignment:}

\begin{itemize}
    \item[\textbf{Description:}] By not contributing internally developed code that relates to a project, a misalignment between the internally and externally developed software may arise. A negative consequence may be that the organization unintentionally ends up on a fork that grows with time and creates unnecessary maintenance and patch-work.
    \item[\textbf{Concern [CaseOrg1,2\&3]:}] Cost of maintaining internal fork. Misalignment between internal and external development may prevent or complicate future contributions.
    \item[\textbf{Example [CaseOrg2]:}] I8 explains it as, \textit{``If you don't share, the community may take off in another direction. Then you will stagnate and come to a place that will be very hard to get back from''}.
    \item[\textbf{Mitigation strategy:}] Keep internal fork of concerned OSS as close as possible to the community's.
\end{itemize}

\subsection{Community-centric Complexities}

\subsubsection{CC13 - External interest:}

\begin{itemize}
    \item[\textbf{Description:}] To contribute a software artifact to an existing OSS community, or to create a new community around it, there needs to be an external interest for the artifact.
    \item[\textbf{Concern [CaseOrg1,2\&3]:}] Risk of contribution not being accepted or community not being established.
    \item[\textbf{Example [CaseOrg1]:}] I3 describes it as \textit{``filling in a gap''}. I5 asks the question, \textit{``Is there a general purpose part of this that I can see multiple teams inside my company taking advantage of?''}. External interest should, therefore, be investigated before proceeding with any contribution. In one example where CaseOrg1 ended up creating a new community, I6 explained, \textit{``everyone needs to do it, and it is not really great agreement over what the right methodologies are''}.
    \item[\textbf{Example [CaseOrg2]:}] For \textit{``existing communities that have been around for long you know, or at least get an indication, if it is going to be an uptake or not [of the contribution]''} (I7). Analyzing stakeholders' agendas within a community can further help. \textit{``What's their strategy, whom do they have playing, what are they trying to get done, what chess moves are they making, and if so, which then informs what we contribute, when we contribute, and how strongly we need to be present''} (I1).
    \item[\textbf{Mitigation strategy:}] Investigate external interest and needs within concerned communities and industries, and consider the generality of the use-case that the software artifact solves.
\end{itemize}

\subsubsection{CC14 - Influence in community:}

\begin{itemize}
    \item[\textbf{Description:}] If the external interest within a community is weak, or if the community is heading in a different direction, it may be important to have influence on the concerned community's RE process in order to create traction and approval for the contribution, but also to be able to steer its development if it is accepted~\cite{linaaker2019community}.
    \item[\textbf{Concern [CaseOrg1,2\&3]:}] Low level of influence in the community may prevent or complicate the contribution of a software artifact.
    \item[\textbf{Example [CaseOrg2]:}] I7 highlights that \textit{``It is important for [CaseOrg2] to pick up a leading role in certain communities that we value as strategically important and as a potential way to get an edge against competitors in those communities''}.
    \item[\textbf{Concern [CaseOrg1,2\&3]:}] Target foundation may require a governance model too open which may render in too large scope and loss of control.
    \item[\textbf{Example [CaseOrg1]:}] I1 explains how CaseOrg1 reasoned about the creation of a separate community behind an internally developed Linux distribution, \textit{``We could have contributed that code to the Linux Foundation or the Apache Software Foundation and have them make it a broader project. But we ended up creating our own foundation and collected all the [relevant stakeholders] together to contribute towards this platform that we created. In a way, it is the right way because we wanted to make sure that it served the need of our [main customers] and it didn't get too broad and become an embedded system that everyone uses... So we want to maintain influence and be in a controlling position, so we are one of three members of the steering committee''}.
    \item[\textbf{Mitigation strategy:}] Build influence needed on concerned community's RE process. If e.g., the current level of influence is deemed not high enough to be able to contribute and steer the software artifact, one option is to create a new community where the community's governance structure can be tailored based on the focal organization's needs.
\end{itemize}

\subsubsection{CC15 - Community health:}

\begin{itemize}
    \item[\textbf{Description:}] Contributing to and engaging actively with a community is one way to support its health, i.e., ability to survive throughout time~\cite{wahyudin2007monitoring}, which is an important aspect if an organization is dependent on a specific OSS project~\cite{linaaker2019community}.
    \item[\textbf{Concern [CaseOrg1,2\&3]:}] Not contributing may have a negative impact on the health of the OSS project.
    \item[\textbf{Example [CaseOrg1]:}] I1 explain CaseOrg1's approach as \textit{``We care about the health of the project, will it die because there are not enough contributors maintaining it? We cannot let that project die because we are using it and we would have to swap out the code and go to something else''}. I1 continues, \textit{``We prefer to use something that is already there, and not reinvents the wheel, but if it is not going to be healthy, then we would want to be there just to create a vibrant community, not for influence, but for health''}.
    \item[\textbf{Mitigation strategy:}] Monitor and analyze the health of concerned OSS communities.
\end{itemize}

\end{appendices}
%\end{small}

\bibliographystyle{plain}

\bibliography{bibliography_master.bib}

\end{document}